\documentclass[fleqn,usenatbib]{mnras}

\usepackage[T1]{fontenc}
\usepackage{ae,aecompl}
\usepackage{multirow}

\usepackage{graphicx}	% Including figure files
\usepackage{amsmath}	% Advanced maths commands
\usepackage{amssymb}	% Extra maths symbols
\usepackage{multicol}    % Multi-column entries in tables
\usepackage{bm}		% Bold maths symbols, including upright Greek
\usepackage{pdflscape}	% Landscape pages

\usepackage[T1]{fontenc}
\usepackage{ae,aecompl}
\usepackage[UKenglish]{babel}

\usepackage{txfonts}

\title[BigMDPL eBOSS Y1Q mocks]{Clustering of quasars in the First Year of the SDSS-IV eBOSS survey: Interpretation and halo occupation distribution}

\author[S. Rodr\'iguez-Torres et al.]{Sergio A. Rodr\'iguez-Torres,$^{1,2,3}$\thanks{email: sergio.rodriguez@uam.es Campus de Excelencia Internacional UAM/CSIC Scholar},   
Johan Comparat,$^{1,3}$\thanks{email: j.comparat@csic.es, Severo Ochoa Fellow} 
 Francisco Prada,$^{1,2,4}$ \newauthor Gustavo Yepes,$^3$ Etienne Burtin,$^5$ Pauline Zarrouk,$^5$ Pierre Laurent,$^5$ ChangHoon Hahn,$^6$\newauthor Peter Behroozi,$^7$ Anatoly Klypin,$^{8,9}$ Ashley Ross,$^{10,11}$ Rita Tojeiro,$^{11}$ Gong-Bo Zhao$^{13,11}$\\
 \\
 $^1$ Instituto de F\'isica Te\'orica, (UAM/CSIC), Universidad Aut\'onoma de Madrid, Cantoblanco, E-28049 Madrid, Spain\\
 $^2$ Campus of International Excellence UAM+CSIC, Cantoblanco, E-28049 Madrid, Spain\\
 $^3$ Departamento de F\'isica Te\'orica M8, Universidad Auton\'oma de Madrid (UAM), Cantoblanco, E-28049, Madrid, Spain\\
 $^4$ Instituto de Astrof\'isica de Andaluc\'ia (CSIC), Glorieta de la Astronom\'ia, E-18080 Granada, Spain\\
 $^5$ CEA, Centre de Saclay, IRFU/SPP, F-91191 Gif-sur-Yvette, France\\
 $^6$ Center for Cosmology and Particle Physics, Department of Physics, New York University, NY 10003, New York, USA\\
 $^7$ Astronomy and Physics Departments and Theoretical Astrophysics Center, University of California, Berkeley, Berkeley CA 94720, USA\\
  $^8$ Astronomy Department, New Mexico State University, Las Cruces, NM, USA\\
 $^9$ Severo Ochoa Associate Researcher at the Instituto de F\'isica Te\'orica (UAM/CSIC), Madrid, Spain\\
$^{10}$ Center for Cosmology and AstroParticle Physics, The Ohio State University, Columbus, OH 43210, USA\\
$^{11}$ Institute of Cosmology \& Gravitation, Dennis Sciama Building, University of Portsmouth, Portsmouth, PO1 3FX, UK\\
 $^{12}$ School of Physics and Astronomy, University of St Andrews, North Haugh, St Andrews KY16 9SS, UK\\
$^{13}$ National Astronomy Observatories, Chinese Academy of Science, Beijing, 100012, P.R.China}

\pubyear{2016}

\begin{document}
%\label{firstpage}
\maketitle

\begin{abstract}
In current and future surveys, quasars play a key role. The new data will extend our knowledge of the Universe as it will be used to better constrain the cosmological model at redshift $z>1$ via baryon acoustic oscillation and redshift space distortion measurements. Here, we present the first clustering study of quasars observed by the extended Baryon Oscillation Spectroscopic Survey. We measure the clustering of $\sim 70,000$ quasars located in the redshift range $0.9<z<2.2$ that cover 1,168 deg$^2$. We model the clustering and produce high-fidelity quasar mock catalogues based on the BigMultiDark Planck simulation. Thus, we use a modified (Sub)Halo Abundance Matching model to account for the specificities of the halo population hosting quasars. We find that quasars are hosted by halos with masses $\sim10^{12.7}M_\odot$ and their bias evolves from 1.54 ($z=1.06$) to 3.15 ($z=1.98$). Using the current eBOSS data, we cannot distinguish between models with different fractions of satellites. The high-fidelity mock light-cones, including properties of halos hosting quasars, are made publicly available.
\end{abstract}


%%%%%%%%%%%%%%%%%%%%%%%%%%%%%%%%%%%%%%%%%%%%%%%%%% 

%%%%%%%%%%%%%%%%% BODY OF PAPER %%%%%%%%%%%%%%%%%%
% \tableofcontents
\section{Introduction}
\label{sec:intro}

How quasars populate the large-scale structure is a puzzle in modern cosmology. It is known that these objects trace the dark matter density field. So, using measurements of the Baryon Acoustic Oscillations (BAO) or redshift space distortions (RSD) from quasars, one can infer information of the cosmological model. However, for these studies or to increase the knowledge of the evolution of quasars, we require a good estimation of their distribution at all scales. Thus, spectroscopic surveys and high-fidelity galaxy mocks from simulations are a great help when solving many riddles concerning quasars.

Large galaxy spectroscopic surveys are an excellent tool to construct a precise 3D map of our Universe. They allow us to study the distribution of different populations in the Universe and constrain cosmological information via BAO scale or RSD measurements. The Sloan Digital Sky Survey \citep[SDSS;][]{york2000} and the two degree field galaxy redshift survey \citep[2dFGRS;][]{norberg2001} first measured the BAO scale in the local universe \citep{eisenstein2005,cole2005}. The Baryon Oscillation Spectroscopic Survey \citep[BOSS;][]{dawson2013}, included in the Sloan Digital Sky Survey III program \citep[SDSS-III][]{eisenstein2011}, recently provided accurate redshifts for 1.5 million galaxies as faint as $i=19.1$, that cover the redshift range $0.2<z<0.75$ on 10,000 square degrees. In combination with SDSS-I/II \citep{york2000}, it provided a sub-percent level measurement of the position of the BAO peak at redshift z=0.57 \citep{alam2016}. SDSS is an example of how spectroscopic surveys can provide strong constraints on our knowledge of the Universe.

Bright quasars constitute the best targets to sample the matter field at high redshift with a small exposure time. Indeed, quasars bear an Active Galactic Nucleus (AGN) that generates light which outshines the entire host galaxy. SDSS I/II published a sample of $\sim100,000$ confirmed quasars \citep{schneider2010} and SDSS-III observed $\sim 170,000$ quasars with redshift $2.1<z<3.5$ as faint as $g=22$ \citep{PARIS2014}. Using both samples, the BAO feature was measured to a few per cent in the Lyman-$\alpha$ (Ly$\alpha$) forest \citep{font-ribera2014,Delubac2015}. Despite the large sample of quasars observed by the SDSS programs, there is still a large region in redshift ($1<z<2.1$) that ought to be studied by targeting quasars fainter than $i=19.1$ in the SDSS imaging. Recent data from other experiments \citep[][e.g., WISE]{wright2010} provides additional information to best target quasars. A cutting-edge target selection algorithm was implemented in \citet{myers2015} and is being observed by the extended Baryon Oscillation Spectroscopic Survey \citep[eBOSS; ][]{dawson2016}, part of the SDSS-IV program. It will increase the number of quasars found by SDSS I/II in the redshift range $0.9<z<2.2$ by a factor of five . This new sample will cover $\sim7,500$ deg$^2$, increasing both the volume and the low number density of the previous samples. It is designed to measure the BAO scale with quasars as tracers of the matter field. In this study, we consider the eBOSS First Year QSO data (hereafter Y1Q). For more details, please see Section \ref{sec:survey}.

Different models have been used to analyse the clustering of quasars. In the literature, many studies focus on the linear regime (large scales). At these scales, correlation function can be described by a power-law \citep[e.g.,][]{Chehade2016}, mostly due to the intrinsic low density of quasars. A more sophisticated method used to model the galaxy clustering and generate mock catalogues is the Halo Occupation Distribution \citep[HOD;][]{jing1998,Peacock2000,Scoccimarro2001,Berlind2002,Cooray2002,zheng2005}. The HOD model recovers the quasar clustering, but its parameters are largely degenerate, producing poor constraints on the host halo masses and satellite fraction \citep{richardson2012,shun2013}. Galaxy samples have also been studied with another method, namely halo abundance matching (HAM), which reproduces the clustering of complete galaxy samples with a reasonable agreement \citep[e.g.,][]{kravtsov2004,conroy2006,behroozi2010,guo2010,trujillo2011,nuza2013,reddick2013}. By including the stellar mass distribution (or luminosity distribution) the HAM also accounts for incomplete samples \citep[e.g.][]{rodrigueztorres2016}. HAM requires knowledge of the stellar mass function, the scatter in the stellar mass to halo mass relation and the incompleteness of the sample. In the case of quasars, obtaining such information is not an easy task. However, modifications of the standard method can be implemented to describe the quasar population.

In the present study, we generate light-cones based on the BigMultiDark Planck simulation \citep[BigMDPL;][]{klypin2014}, using a modified HAM technique to reproduce the Y1Q clustering properties. The BigMDPL is an $N$-body simulation with box size 2.5 $h^{-1}$ Gpc and 3840$^3$ particles, which yields a volume large enough to encompass Y1Q. A variety of mocks, which model different populations of galaxies, have already been constructed using the BigMDPL simulation. They predict, with a good agreement, the observed 2-point and 3-point statistics \citep{favole2015b,guo2015,rodrigueztorres2016}. 

This paper is structured as follows. In Section \ref{sec:data}, we describe the data used in our analysis. Section \ref{sec:cluster} presents the different steps for construction the BigMDPL eBOSS quasar mocks, including how we populate dark-matter halos using a modified HAM algorithm. A set of predictions from our model is shown in Section \ref{sec:results}. Subsequently, we discuss and summarise the most relevant results in Section \ref{sec:disc} and Section \ref{sec:summary}. In this paper, we assume a fiducial $\Lambda$CDM cosmology with the \textsc{Planck-i} parameters $\Omega_m=0.307$, $\Omega_B=0.048$, $\Omega_\Lambda=0.693$ \citep{planck2014}.

\section{Data}
\label{sec:data}

\subsection{eBOSS QSO survey and clustering}
\label{sec:survey}

The extended Baryon Oscillation Spectroscopic Survey \citep{dawson2016} is part of a six year SDSS-IV program (fall 2014 to spring 2020). It combines the potential of SDSS-III/BOSS and new photometric information to optimise target selection and extend BAO studies to higher redshift. eBOSS uses the 2.5-meter Sloan Foundation Telescope at Apache Point Observatory \citep{gunn2006} and the same fiber-fed optical spectrograph as BOSS, where each fiber subtends a $2^{\prime\prime}$ diameter of the sky \citep{smee2013}. This survey will provide redshifts for $300,000$ Luminous Red Galaxies (LRG) in the redshift range $0.6<z<1.0$, a new sample of $\sim200,000$ Emission Line Galaxies (ELG) at redshift $z>0.6$, more than $500,000$ spectroscopically-confirmed quasars at $0.9<z<2.2$ and $\sim120,000$ new Ly$\alpha$ forest quasars at redshift $z>2.1$.

eBOSS dedicates 1,800 plates to cover an area of 9,000 deg$^2$: 1,500 plates to measure LRG and QSO redshifts on 7,500 deg$^2$ and 300 plates to measure ELG redshifts on 1,000 deg$^2$. The first two years, observations were dedicated to the QSO and LRG samples. In order to maximise the tiling completeness and fiber efficiency in the LRG/QSO sample, a tiered-priority is adopted \citep{dawson2016}, where the QSO targets have maximal priority and are assigned to fibers first.

eBOSS has adopted two approaches to target quasars for redshift $>0.9$ \citep{myers2015}. In the first approach``Clustering'' quasar targets (\textsc{qso\_core}) are used as a direct tracer of the large-scale structure in the redshift range $0.9<z<2.2$. The second approach consists in detecting quasars at $z>2.1$ to map the large-scale structure via absorption of the Ly$\alpha$ forest \citep{palenque2016}. 
\begin{enumerate}
\item The \textsc{core} quasar sample is constructed combining optical selection in $ugriz$ using a likelihood-based routine called \textsc{XDQSOz} \citep{bovy2011}, with a midIR-optical colour-cut. eBOSS \textsc{core} selection (to $g<22$ OR $r < 22$) should obtains $\sim$ 70 quasars per sq. deg. at redshifts $0.9 < z < 2.2$ and about 7 quasars deg$^{-2}$ at $z > 2.2$.
\item The Ly$\alpha$ quasar selection is based on variability in multi-epoch imaging from the Palomar Transient Factory \citep{palenque2016}. It recovers an additional 3 or 4 quasars deg$^{-2}$ at $z>2.2$ to $g<22.5$. A linear model of how imaging systematics affect target density recovers the angular distribution of eBOSS \textsc{core} quasars over 96.7\% (76.7\%) of the SDSS North (South) Galactic Cap area \citep{myers2015}.
\end{enumerate}

\citet{busca2013} measure the BAO scale using Ly$\alpha$ quasars from the BOSS data. \citet{font-ribera2014} also give measurements of this scale using the cross-correlation between visually confirmed quasars with the Ly$\alpha$ forest absorption. One of the goals of eBOSS is to provide a first detection of the BAO scale using only the \textsc{core} quasar sample. 

In this context, we focus our study on the spectroscopically confirmed QSO using the Y1Q data which includes 68,269 objects that cover 1,168 deg$^2$ of the sky. Table \ref{table:nbar} shows the abundance of \textsc{core} QSO at different redshift ranges.
\begin{table}
\centering
\caption{Distribution of the Y1Q sample in four redshift bins. $\bar{n}$ represents the comoving number density of QSO, $N$ is the number of QSO and $V$ is the comoving volume of the redshift bin subtended by 1168 deg$^2$. The last line shows the values for the complete sample.}
\label{table:nbar}
\begin{tabular}{cccc} \hline \hline
 \multirow{2}{*}{Redshift}& $\bar{n}$ & \multirow{2}{*}{$N$} & $V$  \\
  & [$10^{-5}$Mpc$^{-3}h^3$] & & [$10^9$ $h^{-3}$Mpc$^3$] \\ \hline 
 $0.9<z<1.2$ & 1.36 & 13,484 & 0.99 \\
 $1.2<z<1.5$ & 1.48 & 17,578 & 1.19\\
 $1.5<z<1.8$ & 1.36 & 17,778 & 1.31\\
 $1.8<z<2.2$ & 1.05 & 19,429 & 1.84\\ \hline 
 $0.9<z<2.2$ & 1.28 & 68,269 & 5.34\\ \hline 
\end{tabular}
\end{table}

\subsection{Redshift error and statistical weights}
\label{sec:redshift_errors}
eBOSS expects a redshift precision better than 300 s$^{-1}$km RMS for the QSO \textsc{core} at $z<1.5$ and better than [300+400(z-1.5)] km s$^{-1}$ at $z>1.5$ \citep{myers2015}. It corresponds to redshift errors of the order of 1$\times10^{-3}$ for $z<1.5$ and $\sim5\times10^{-3}$ for larger redshift. These errors have an important impact on scales smaller than 10 $h^{-1}$Mpc (see Appendix \ref{app:errors}). For this reason we add redshift errors to the mock catalogues using these upper limits. In addition, less than 1\% of the sample is expected to have catastrophic redshift errors.

In order to include the observed redshift precision in the light-cones, we model redshift errors using a Gaussian distribution with mean value $z_{true}$ and width $\Delta z$,
\begin{equation}
z=z_{true}+\Delta z\mathcal{N}(0,1),
\end{equation}
where $\mathcal{N}(0,1)$ is a random number coming from a Gaussian distribution with mean 0 and standard deviation 1 and
\begin{equation}
\label{eq:rdserr}
\Delta z = \begin{cases}
300\text{ km s}^{-1}c^{-1} & \textrm{if}\; z<1.5\\
[300+400(z-1.5)]\text{km s}^{-1}c^{-1} & \textrm{if}\; z\geq1.5,
\end{cases}
\end{equation}
 $c$ represents the speed of light. We also include 1\% of catastrophic redshift errors, which introduces a reduction in the amplitud of the correlation function of $\sim$ 1\% at all scales (Appendix \ref{app:errors}). In order to include these errors, we randomly select 1\% of the mock galaxies and replace their redshift by a random value within the range of the catalogue.

A correct estimation of redshift errors is important in order to understand the behaviour of the clustering at small scales. The monopole of the correlation function is affected by over 50\% at scales below 10 $h^{-1}$Mpc. The impact is larger on the quadrupole, where the effects are detected at scales below 40 $h^{-1}$Mpc \citep{reid2011}. In Appendix \ref{app:errors}, we explore with more detail the impact of these errors on clustering measurements. Nevertheless, even if we model the redshift errors, this is still an approximation which can introduce unphysical effects. This can result in a wrong estimation of the model's parameters if scales affected by errors are included in the fitting procedure. For this reason, we fix the parameters using the monopole of the correlation function between 10 and 40 $h^{-1}$Mpc, where the impact of redshift errors decreases and the effects of the cosmic variance and shot noise become smaller (Appendix \ref{app:errors}).

In addition to redshift measurement, the 5-$\sigma$ detection limit for point sources (also called depth) of the SDSS photometric survey varies across the footprint and differs for each band. The amplitude of the variations implies that faint targets end up very close to the detection limit. These targets are then more likely to be missed by the target selection algorithm. eBOSS corrects this effect by applying a depth-dependent weight, called ``systematics weight'' $w_{sys}$ to each quasar (see Laurent et al. in prep. for a detailed description).

Finally, eBOSS takes fiber collisions and redshift failures into account by using using weights for each, $w_{cp}$ and $w_{zf}$ respectively. Those quantities are initialised to one for all objects. Then, if a quasar has a nearest neighbour with a redshift failure or its redshift was not obtained because it was in a close pair, $w_{zf}$ or $w_{cp}$ are increased by one \citep{ross2012}. Including all these effects, the total weight for each quasar in the observed data is given by
\begin{equation}
w_{\textsc{q}}=w_{\textsc{fkp}}w_{sys}(w_{cp}+w_{zf}-1),
\end{equation}
where $w_{\textsc{fkp}}$ is the density weight applied for an optimal estimation of the 2-pt function and is defined by the expression \citep{feldman1994}
\begin{equation}
  \label{eq:fkp}
 w_{\textsc{fkp}}=\frac{1}{1+n(z)P_{\textsc{fkp}}},
\end{equation}
where $n(z)$ is the number density at redshift $z$ and $P_{\textsc{fkp}}=6000$ $h^{-3}$ Mpc$^3$. 

Corrections for fiber collisions using close pair weights do not provide an accurate clustering signal at small scales \citep{guo2012,hahn2016}. However, in the quasar sample the distribution of objects is disperse and the number of collided pairs is very small. Additionally, our analysis does not use scales below 10 $h^{-1}Mpc$, so the close pair correction is good enough for our purpose. In the case of the simulated quasars, we include FKP weights but do not simulate the effects that require any of the additional weights applied to the data sample.

\subsection{The eBOSS BigMultiDark light-cone}
\label{sec:simu}

The suite of MultiDark\footnote{\url{http://www.multidark.org/}} Planck simulations adopts a flat $\Lambda$CDM model with \textsc{Planck-I} cosmological parameters \citep{planck2014}: $\Omega_m=0.307$, $\Omega_B=0.048$, $\Omega_\Lambda=0.693$, $\sigma_8=0.829$, $n_s=0.96$ and a dimensionless Hubble parameter $h=0.678$. We only use two of the $N$-body simulations described in \citet{klypin2014}. The BigMultiDark (BigMDPL) has a box length of 2.5 $h^{-1}$ Gpc with 3840$^3$ particles of mass $2.4\times10^{10}$ $h^{-1}M_{\odot}$ and the MultiDark Planck (MDPL) has a box length of 1.0 $h^{-1}$ Gpc with 3840$^3$ particles with a mass of $1.5\times10^{9}$ $h^{-1}$ $M_{\odot}$. Both were built with \textsc{GADGET-2} \citep{springel2005} using initial Gaussian fluctuations generated with the Zel'dovich approximation at redshift $100$. 

From the dark matter catalogues of the simulation, halos are defined with the Robust Overdensity Calculation using K-Space Topologically Adaptive Refinement halo finder \citep[\textsc{RockStar};][]{behroozi2013}. Spherical dark matter halos and sub-halos are identified using an approach based on adaptive hierarchical refinement of friends-of-friends groups in six-phase space dimensions and one time dimension. \textsc{RockStar} computes halo mass using spherical overdensities of a virial structure \citep{bryan1998}. Before calculating halo masses and circular velocities, the halo finder performs a procedure which removes unbound particles from the final mass of the halo \footnote{\url{http://www.cosmosim.org/}}. We include observational effects and construct a catalogue with similar volume to the eBOSS sample, by making light-cones based on different snapshots of the BigMDPL simulation.

We perform the modified halo abundance matching by using the maximum circular velocity of the halo ($V_{max}$) in order to link dark matter halos and quasars. The maximum circular velocity is one of the best candidates for matching dark matter halos and galaxies \citep{reddick2013}. $V_{max}$ can be related to the virial mass of the halo through a power-law given by 
\begin{equation}
V_{max}= \beta(z)[M_{vir}E(z)/(10^{12}h^{-1}\text{Mpc})]^{\alpha(z)},
\end{equation}
where, $E(z)=\sqrt{\Omega_{\Lambda,0}+\Omega_{m,0}(1+z)^3}$, $\log_{10}\beta(z)=2.209+0.060a-0.021a^2$ and $\alpha(z)=0.346-0.059a+0.025a^2$, with $a=1/(1+z)$ the scale factor \citep[see][]{rodriguezpuebla2016}. There are better candidates to perform the matching between dark matter halos and galaxies, such as, the maximum circular velocity along the whole history of the halo ($V_{peak}$). However, the BigMDPL simulation has a small number of snapshots (4) in the quasar redshift range thus preventing a good estimation of quantities that are computed by tracing halos between snapshots. For this reason, we use $V_{max}$ to implement our model. Differences between $V_{peak}$ and $V_{max}$ become important in case of substructures, while the selection of host halos is similar with both quantities. \cite{reddick2013} show a significantly larger amount of subhalos when $V_{peak}$ is used rather than other quantities. However, in our model the impact of choosing $V_{max}$ can be compensated by using the fraction of satellites as a free parameter. Furthermore, the poor information of the one halo term in the quasar sample and the large errors in observations will not allow us to distinguish which quantity performs the matching better.

Table \ref{table:multidark} presents the deviation of each simulation from a model of the complete mass function \citep{comparat2017}, which is obtained by fitting a data set that contains the complete part of each of the MultiDark Planck simulation (SMDPL, MDPL, BigMDPL, HMDPL). Masses in Table \ref{table:multidark} fulfil the condition given by
  \begin{equation}
    \label{eq:mass}
    N_{sim}(M_{200}>M_i)/N_{mod}(M_{200}>M_i)<percentage,
  \end{equation}
where $N_{sim}$ is the number of objects in the simulation with $M_{200}$ smaller than the threshold mass $M_i$ and $N_{mod}$ is the corresponding number of halos in the model. Previous works showed that quasars live in halos with masses of the order of $\log(M/M_\odot)\sim12.5$ \citep{shun2013,Chehade2016}. Both simulations mentioned above are complete for this mass as is shown in Table \ref{table:multidark}. But depending on the dispersion of the distribution of halos hosting QSO, a small fraction of halos coming from the incomplete part of the simulation enter in the final mock. We quantify the effect of the resolution in our catalogues with the MDPL, where this effect is negligible thanks to its higher resolution. MDPL has enough resolution to cover the halo mass range for the QSO population. However, its volume is smaller than the one covered by eBOSS, so one cannot construct a complete light-cone without box replications. Furthermore the shot noise from a mock using this volume is very large, due to the low number density of the observed sample. In Appendix \ref{app:resolution}, we show this effect by comparing the mocks generated from both simulations.
\begin{table*}
\centering
 \caption{Deviation from the mass function at redshift 0 for the MDPL and the BigMDPL simulations. The masses and the maximum circular velocities are the threshold above which the completeness in this box relative to the mass function is higher than the percentage given in the header (see Equation \eqref{eq:mass}). The corresponding number of particles is provided in brackets.}
 \label{table:multidark}
 \begin{tabular}{ccccccccc}\hline \hline
  & \multicolumn{4}{c}{$\log(M_{200c}(z)/M_\odot)$} & 
      \multicolumn{4}{c}{$V_{max}$} \\
  fraction & 80\% & 90\% & 95\% & 97\% & 80\% & 90\% & 95\% & 97\% \\ \hline 
   & \multicolumn{8}{c}{central halos} \\ \hline 
MDPL & 11.04 (71) & 11.10 (82) & 11.26 (119) & 11.61 (266) & 57.3 & 68.6 & 98.3 & 121.9\\ 
BigMD & 12.22 (69) & 12.28 (79) & 12.32 (87) & 12.36 (98) & 131.0 & 145.9 & 201.6 & 299.3\\ \hline 
% & \multicolumn{8}{c}{satellite halos} \\ \hline 
%MDPL & 11.065 (76) & 11.125 (88) & 11.255 (119) & 11.565 (243) & 80.09 & 83.61 & 91.80 & 114.70\\ \hline 
\end{tabular}
\end{table*}

We include the redshift evolution in the number density and of the clustering when constructing light-cones from the BigMDPL simulation. These light-cones cover the redshift range $0.9<z<2.2$ and 1,481.75 deg$^2$ of the sky, which is comparable with the area of Y1Q. The mocks are built with the SUrvey GenerAtoR code \citep[\textsc{SUGAR};][]{rodrigueztorres2016}. In this procedure, we use all available snapshots from the BigMDPL simulation, $z=$2.145, 1.445,1, 0.8868. In order to analyse the effects of the incompleteness, we select only the closest snapshots from the MultiDark simulation ($z=$1.425, 0.987, see Appendix \ref{app:resolution}). We present results from three different light-cones, the first one uses a single set of parameters to describe the Y1Q (BigMDPL-QSO). The second one is obtained by fitting the clustering in four redshift bins with a different set of parameters (BigMDPL-QSOZ). The last light-cone uses a single set of parameters, but only host halos are included (the fraction of substructures is equal to zero, BigMDPL-QSO-NSAT).

\subsection{Galaxy Mocks for QSO (GLAM)}

In order to estimate the uncertainties in the clustering measurements, we use the GaLAxy Mocks (GLAM) scheme for the eBOSS quasar sample. For this application, GLAM implements a new parallel particle mesh method \citep[PPM-GLAM;][]{klypin2017} to construct the dark matter density field and an optimisation to populate the simulation with quasars (Comparat et al., in prep). We run the \textsc{SUGAR}-code to construct light-cones \citep{rodrigueztorres2016}. Errors are extracted from the covariance matrix of 1000 GLAM-QSO mocks which cover the same area as the data. They are computed using the diagonal terms, $\sigma_i(x_i)=\sqrt{C_{ii}}$, thus these errors correspond to one standard deviation ($1\sigma$) away from the mean value of the mocks. We use the covariance matrix estimator given by 
  \begin{equation}
    \label{eq:covmat}
    C_{ij}=\frac{1}{n_s-1}\sum\limits_{k=1}^{n_s}\big(x_i^k-\mu_i\big)\big(x_j^k-\mu_j\big),
  \end{equation}
where $n_s$ is the total number of mocks and the mean of each measurement is
\begin{equation}
  \label{eq:mean}
  \mu_i=\frac{1}{n_s}\sum\limits_{k=0}^{n_s}x_i^k.
\end{equation}
Using the covariance matrix from these mocks we perform the fitting with the $\chi^2$ statistics,
\begin{equation}
  \label{eq:chi2}
  \chi^2=\sum\limits_{ij}\big[x_i^d-x_i^m\big]C_{ij}^{-1}\big[x_j^d-x_j^m\big],
\end{equation}
where $x_ i^m$ and $x_i^d$ are the measurements from the model and the data in the bin $i$ respectively. $\chi^2$ values presented in this work are computed from the monopole of the correlation function.

\section{Clustering Model}
\label{sec:cluster}

One of the best ways to study the observed clustering of a survey is to simulate not only the effect of the gravity on the dark matter but also on the baryonic matter. In this case, stellar physics should be included to provide a direct prediction of the relation between dark matter halos and the galaxies and their evolution in time. This approach is undertaken by hydrodynamical simulations, that include galaxy formation processes, stellar physics and AGN feedback. EAGLE \citep{rahmati2015} and ILLUSTRIS \citep{sijacki2015} are two of the most recent realisations which predict a realistic distribution of galaxies and quasar populations. However, these simulations are constructed in rather small boxes of $\sim 75 h^{-1}$ Mpc and this impedes studies of the large scale structure. The large amount of computational resources required for a hydrodynamic simulation is prohibitive and the computation of volumes comparable to observations nearly infeasible. 

An alternative approach, cheaper in computational time, is to use the dark matter only simulations and add galaxies in a statistical way. There are two widely used models based on these statistical relations. The first one is the Halo Occupation Distribution \citep[HOD; e.g.,][]{guo2014}, which gives the probability, $P(N|M_h)$, that a halo of mass $M_h$ hosts $N$ galaxies. This probability is described by a fitting formula, which is fixed using the clustering measurements from the observational data. The second method to populate the dark matter halos is the Halo Abundance Matching \citep[HAM; e.g.,][]{reddick2013}. This model assumes that the most massive galaxies populate the most massive halos. 

\subsection{The modified SHAM model}
\label{sec:model}

\citet{favole2015b} introduced a modified (Sub)Halo Abundance Matching (SHAM), designed to reproduce the clustering of the BOSS ELG sample. They select halos from the simulation using a probability function which is the sum of two terms corresponding to host and satellite halos. This probability is a Gaussian function described by three parameters: the mean mass, the width of the distribution and the satellite fraction. This method is useful to describe incomplete samples, such as the Y1Q, which is not complete in halo mass or stellar mass whatsoever. In this paper, we use a similar model to study the clustering of quasars. \citet{favole2015b} use the virial mass of halos to implement their method. Instead of that, we use $V_{max}$ and assume that the distribution of halos hosting quasars has a Gaussian shape. The most general model is split in central and satellite halos as done in \citet{favole2015b}. When a QSO is located in the centre of a host halo, it is denoted as a central QSO. The satellite fraction refers to the fraction of QSO living in a sub-halo. This fraction does not represent systems of binary quasars. The central halo which is the counterpart of a satellite QSO can host another kind of galaxy.  

In the case of quasars, we do not use the luminosity or the stellar mass of the observed sample. Our model only uses the $V_{max}$ distribution of halos, as done by \citet{nuza2013}. \citet{rodrigueztorres2016} extend the HAM technique implemented by \citet{nuza2013} using the stellar mass function and modelling the incompleteness of the sample. In that study, galaxies are assigned to halos via a standard HAM and then they are downsampled to obtain the observed stellar mass distribution. Here, we assume that the intrinsic scatter between quasars and dark matter halos, plus the incompleteness of the sample will produce a $V_{max}$ distribution with a Gaussian shape. Then, the model orders halos by $V_{max}$ and downsamples objects as done by \citet{rodrigueztorres2016}.

\subsection{Implementation}

Assuming that the final $V_{max}$ distribution of the simulated quasar catalogue is Gaussian, we need to construct a probability distribution function which selects halos from the complete simulation based on this condition. In a general case, the $V_{max}$ distribution of the final catalogue will be
  \begin{align*}
    \phi_{\textsc{qso}}(V_{max})=&\phi_{\textsc{qso}}^{s}+\phi_{\textsc{qso}}^{c}\\
    =&P_s(V_{max}) \phi_{sim}^{s}(V_{max})+P_c(V_{max}) \phi_{sim}^{c}(V_{max})\\
    =&\mathcal{G}_{s}(V_{max})+\mathcal{G}_{c}(V_{max}),
  \end{align*}
where $\phi_{sim}^c$ and $\phi_{sim}^s$ represent the $V_{max}$ distribution of host halos and subhalos respectively, $\mathcal{G}_c$ and $\mathcal{G}_s$ are Gaussian functions with mean $V_{mean}$, standard deviation $\sigma_{max}$ and each one is normalised using
\begin{equation*}
    \int\mathcal{G}_{s}(V_{max},z)dV_{max}=N_{tot}(z)f_{sat}
\end{equation*}
\begin{equation*}
    \int\mathcal{G}_{c}(V_{max},z)dV_{max}=N_{tot}(z)(1-f_{sat}) 
\end{equation*}
where $N_{tot}(z)$ is the total number of quasars per redshift bin given by the observed number density.

In order to construct the probability distribution, we sort all halos in the simulation and compute the maximum circular velocity function $(V_{max})$ for sub-halos and host halos separately. Using the fraction of satellites as a free parameter and the observed number density, we normalise the Gaussian distribution for central and satellite halos. We split all halos of the simulation in bins of $V_{max}$ and compute the probability of assigning a quasar to a dark matter halo (central or satellite) per bin as 
\begin{equation}
 \label{eq:prob}
 P_{s/c}(V_{max})=\frac{N^{gaus}_{s/c}}{N^{tot}_{sub/host}},
\end{equation}
where $N^{tot}_{sub/host}$ is the total number of sub/host halos in the range [$V_{max}-\Delta V_{max}/2$, $V_{max}+\Delta V_{max}/2$] and $N^{gaus}_{s/c}$ is the number of satellite/central quasars necessary to produce the final Gaussian shape. Using equation \eqref{eq:prob}, we downsample all halos in the simulation to obtain the QSO mock catalogue.

Our model consists of 5 different parameters, the mean and standard deviation values for satellite and central distributions and the fraction of satellites. However, we assume the same mean and standard deviation for central and satellite quasars thus decreasing the number of parameters. In addition, the current data does not provide enough information at small scales ($<1.0$ $h^{-1}$Mpc) to extract precise information about the standard deviation of the distribution and the satellite fraction of the eBOSS QSO sample. For these reasons, our unique parameter to fit the clustering is the mean value of the distribution ($V_{mean}$). 

\subsection{Parameters}
\label{sec:parameters}

The most general model is defined by three parameters. However, due to the poor information at small scales, we only use one free parameter ($V_{mean}$) to describe the Y1Q sample. Figure \ref{fig:chi2} presents the $\chi^2$ maps we obtain for different combinations of the three parameters $V_{mean}$, $\sigma_{max}$ and $f_{sat}$. We find the satellite fraction, $f_{sat}$, to be degenerate with $V_{mean}$ (Left-hand panel Figure \ref{fig:chi2}) and this degeneracy could be broken only with information from the one halo term. However, the current Y1Q data does not allow going to those scales. For this reason, we do not fix the number of satellites in two of the three mocks presented, which means that host halos and subhalos are not distinguished when the selection is implemented. In addition, just as \citet{favole2015b}, we do not find a dependency of the clustering with the width of the Gaussian distribution ($\sigma_{max}$). $\sigma_{max}$ cannot be constrained with the current data as is shown in the right-hand panel of Figure \ref{fig:chi2}. In the mass regime where QSOs live, $\sigma_{max}$ impacts the clustering at small scales ($<0.5$ $h^{-1}$ Mpc), so it is not possible to constrain this parameter. 

In the case of quasars, at scales larger than 1.0 $h^{-1}$ Mpc, the clustering amplitude only depends on $V_{mean}$. In order to fix $\sigma_{max}$, we use previous results in the literature. The model shown in \citet{Chehade2016} is consistent with a width in $V_{max}$ of $\sigma_{max}=45$ km $s^{-1}$. However, due to the resolution of BigMultiDark, we decrease this value to $\sigma_{max}=30$ km $s^{-1}$. If we use larger values of $\sigma_{max}$, we will include a larger fraction of halos from the incomplete mass region of the simulation. Fixing $\sigma_{max}=30$ km $s^{-1}$, we ensure that the BigMDP light-cones have only $\sim$2\% of halos selected from regions where the incompleteness is greater than 10. Thus, we avoid including any unphysical effects coming from the low resolution of the simulation.

Thus, our model describes the quasar sample with a single parameter which is fixed by minimising the $\chi^2$ distribution. As mentioned previously, we use the monopole of the correlation function between 10 and 40 $h^{-1}$ Mpc (10 data points shown in Figure \ref{fig:monoevo}), thereby avoiding systematic effects that influence the clustering measurements at small scales. Varying $V_{max}$, we find that the $\chi^2$ distribution is well described by a quadratic function. This is used to find the parameter which best represents the data.

\section{Results}
\label{sec:results}

\begin{figure}
 \includegraphics[width=86mm]{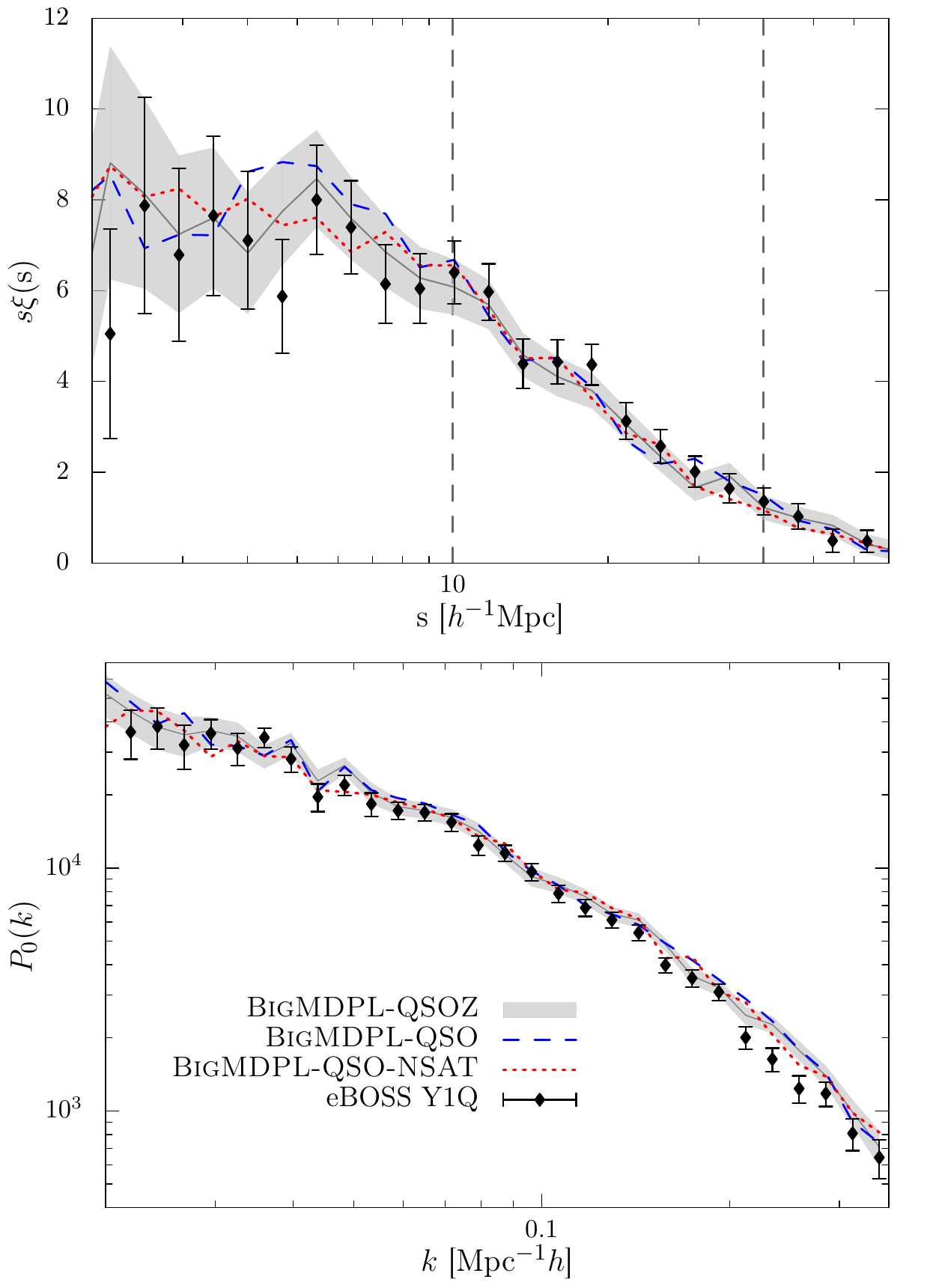}
 \caption{\textit{Top panel:} Monopole of the correlation function in configuration space of Y1Q (points with error bars). The shaded area represents the BigMDPL-QSOZ light-cone fitted in four different redshift bins. The dashed line represents the BigMDPL-QSO light-cone fitted on a single redshift bin and the dotted line is the BigMDPL-QSO-NSAT. The vertical lines represent the limit values used for fitting the parameters. \textit{Bottom panel:} Monopole of power spectrum of the Y1Q (points with error bars) and the three BigMDPL light-cone. The agreement between the best model and the data is remarkable. Error bars and dashed areas are computed using 1000 GLAM catalogues and correspond to $1-\sigma$ deviation from the mean value}. Differences at high $k$ are due to redshift errors.
 \label{fig:monoevo}
\end{figure}

We compare the Y1Q 2-point correlation function (2PCF) with that of the mocks using the $\chi^2$ statistics with 9 degrees of freedom (10 data points and 1 parameter). In order to compute the 2PCF we use a modified version of the Correlation Utilities and Two-point Estimation code \citep[\textsc{cute};][]{alonso2012}. We first analyse the complete sample, using the clustering measurements in the redshift range $0.9<z<2.2$. We find the best value for the parameter $V_{mean}=341.2$ km s$^{-1}$, which corresponds to a sample of mock QSO with mean mass $\log[M_{200}/M_\odot] = 12.66\pm0.16$. Figure \ref{fig:monoevo} presents the clustering measurements (2PCF and power spectrum) along with the prediction of the best-fit mock light-cone. We find an excellent agreement between the data and the model for the studied scales. 

When fitting is performed using the clustering of the complete redshift range, the evolution of the mass distribution is not taken into account. In order to investigate this effect, we divide the sample in four redshift bins and find the best parameter to match the clustering in each individual redshift range. It slightly improves the quality of the fits, presented in Table \ref{table:meanVpeak} which gives the best fit values of $V_{mean}$ and their corresponding reduced $\chi^2$. 

\begin{table}
\centering
\caption{Results of the fit per redshift bin. $A$ gives the area in deg$^2$ subtended by the mock light-cone. $z$ bin gives the lower and upper boundary of the redshift bin. $V_{mean}$ is the best fit parameter found. $\log_{10}(M_{200}/M_\odot)$ is the corresponding mean $\pm$ standard deviation of the halo mass of the population selected. $\chi_r^2$ is the reduced $\chi^2$ per 9 degrees of freedom. We fixed $\sigma_{max}=30$ km $s^{-1}$ and $f_{sat}$ is percentage of satellites in the catalogue.}
\label{table:meanVpeak}
\begin{tabular}{cccccc} \hline \hline
 $A$ (deg$^2$) & $z$ bin & $V_{mean}$  (s$^{-1}$km) & $\log_{10}\frac{M_{200}}{M_\odot}$ & $\chi_r^2$ &$f_{sat}$ \\ \hline \hline
\multicolumn{5}{c}{BigMDPL-QSO  \vspace{0.1cm}}\\ %\hline 
1,481.75 & $0.9 - 2.2$ & 341.2$\pm$30.0 & 12.66$\pm$0.16 & 1.78 & 5.3\\ \hline 
\multicolumn{5}{c}{BigMDPL-QSOZ \vspace{0.1cm}}\\ %\hline
 3,275.06 & $0.9 - 1.2$ & 282.8$\pm$30.2 & 12.53$\pm$0.17 & 1.47 & 9.0\\
 2,371.81 & $1.2 - 1.5$ & 324.1$\pm$30.1 & 12.63$\pm$0.14 & 1.85 & 5.0\\
 1,879.13 & $1.5 - 1.8$ & 339.5$\pm$29.9 & 12.69$\pm$0.14 & 1.70 & 4.3\\
 1,481.75 & $1.8 - 2.2$ & 353.5$\pm$29.7 & 12.60$\pm$0.13 & 2.24 & 3.3\\ \hline
\multicolumn{5}{c}{BigMDPL-QSO-NSAT \vspace{0.1cm}}\\ %\hline 
1,481.75 & $0.9 - 2.2$ & 349.5$\pm$30.3 & 12.70$\pm$0.16 & 1.52 & 0.0\\ \hline  
\end{tabular}
\end{table}

Comparing the values of $M_{200}$ presented in Table \ref{table:meanVpeak} with those of Table \ref{table:multidark}, we infer that the best fit mocks have less than 1\% of objects taken from a bin where the completeness is lower than 90\%. The effect of the resolution on the clustering is discussed in more detail in Appendix \ref{app:resolution}.

\begin{figure}
 \includegraphics[width=86mm]{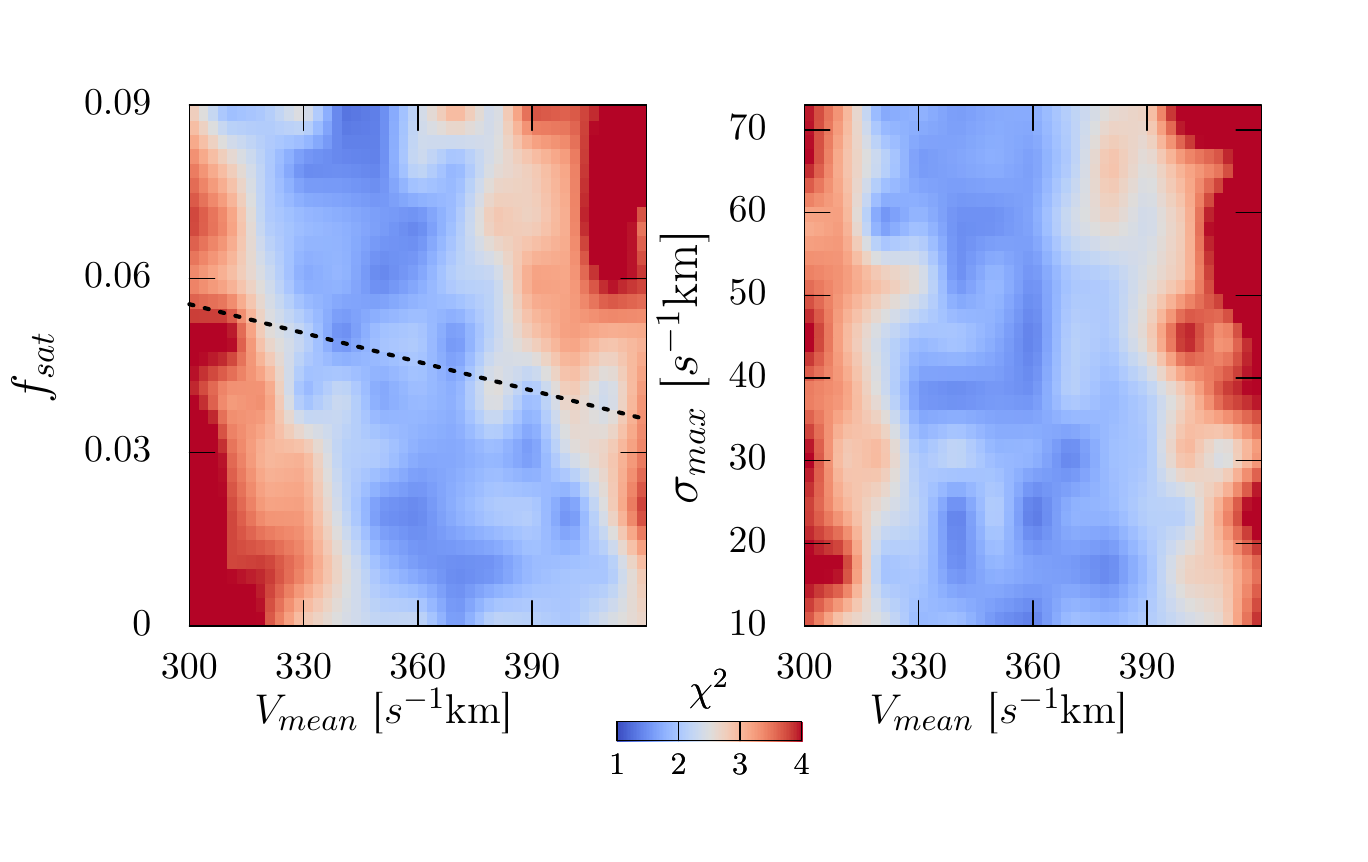}
 \caption{$\chi^2$ maps for the three parameters of the model implemented on the BigMDPL-QSO. The left-hand panel shows the satellite fraction vs. $V_{mean}$. It is possible to note a degeneracy between both parameters. This is why we use the $f_{sat}$ given by the simulation. The Dashed line shows the satellite fraction given by the simulation for different values of $V_{mean}$. The right-hand panel presents $\sigma_{max}$ vs. $V_{mean}$. $\sigma_{max}$ cannot be constrained using the current data.}
 \label{fig:chi2}
\end{figure}

Table \ref{table:meanVpeak} shows the values of satellite fractions of the BigMDPL light-cones. As we explained in Section \ref{sec:parameters}, we do not use $f_{sat}$ as a parameter so the fraction of satellites in the mock has the same dependency with $V_{max}$ as the complete simulation. The third light-cone is the only catalogue where we fix $f_{sat}=0$. We include it to show the impact of removing all substructures from our analysis. The second parameter of the model, $\sigma_{max}$ is also not constrained (see Figure \ref{fig:chi2}). A similar problem was found by \citet{shun2013}, their HOD parameters are largely degenerate and the fraction of satellites is not well constrained. For these reasons, we only vary the mean value of the Gaussian distribution ($V_{mean}$) to fix the clustering of the model.

\subsection{Trends of the QSO clustering with redshift}

\begin{figure}
 \includegraphics[width=84mm]{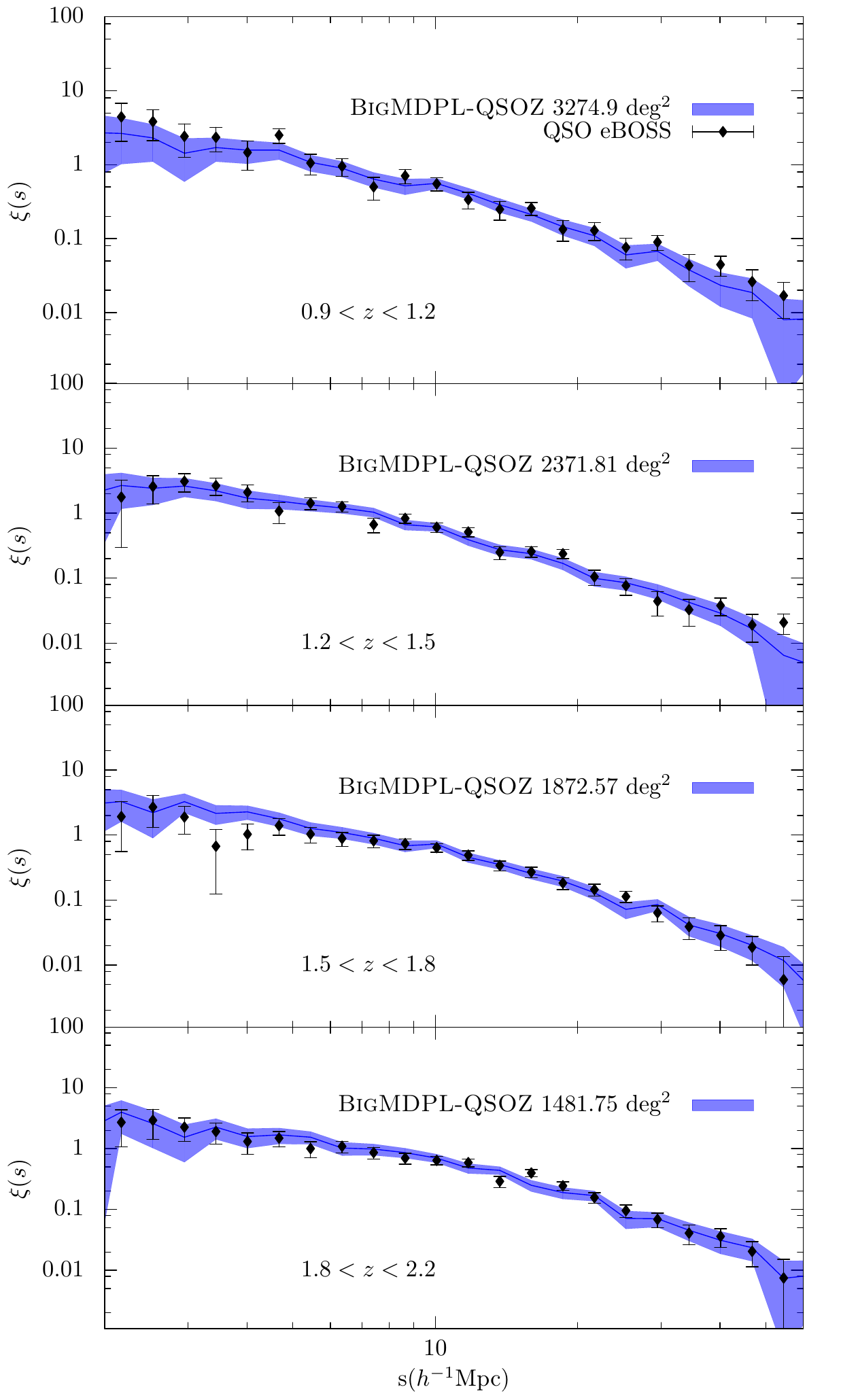}
 \caption{Monopole 2PCF vs. redshift. We show the Y1Q  (points) and the best-fit mock (shaded area) of the BigMDPL-QSOZ light-cone (see Table \ref{table:meanVpeak}). Each panel corresponds to a different redshift bin. Error bars and dashed areas are computed using 1000 GLAM catalogues and correspond to $1-\sigma$ deviation from the mean value.}
 \label{fig:mono}
\end{figure}

The signal of the quasar clustering does not have an important evolution, as shown in Figure \ref{fig:mono}. The monopole varies mildly in the linear regime in all four redshift bins. If we assume a constant distribution of $V_{max}$ for the whole redshift range, the evolution of the dark matter field will produce a non-constant signal of clustering in the different redshifts. In order to reproduce the observed evolution and predict a most realistic linear bias, we divide the complete redshift range into four regions, fitting the clustering of the light-cone in each bin. Table \ref{table:meanVpeak} presents the redshift range and the best-fit parameters found to match the observed data. We use different areas for each redshift bin to maximise the volume used from the simulation. These larger areas increase the statistics and reduce the shot-noise in the 2PCF of the mocks as seen in Table \ref{table:meanVpeak}.

Figure \ref{fig:monoevo} shows the monopole of the correlation function and the power spectrum of the three different mocks (BigMDPL-QSO/QSOZ/QSO-NSAT) compared to the observed data for the whole redshift range. All light-cones can reproduce the eBOSS data with a good agreement. We underline that the BigMDPL light-cones have shot-noise and cosmic variance similar to the data. Due to these large errors in the model and the data, it is difficult to distinguish which light-cone reproduces the data better in the complete redshift range. However, if the model reproduces the clustering at different redshifts, we can estimate the evolution of the bias with better accuracy. 

In order to quantify the difference between two models, we compare them using the Bayes factor. We can compute it with the maximum likelihood
  \begin{equation}
    \label{eq:likeli}
    P(\textbf{x}|\textbf{p})=\frac{|\textbf{\~C}^{-1}|}{(2\pi)^p}\exp\Big[-\frac{1}{2}\sum\limits_{ij}\big(x_i^d-x_i(\text{p})\big) \textit{\~C}^{-1}_{ij}\big(x_j^d-x_j(\text{p})\big)\Big]
  \end{equation}
where x$^d$ represents the data and x(p) the model. We estimate the inverse covariance matrix using equation \eqref{eq:covmat} and correcting for bias using the Hartlap factor \citep{hartlap2007}
  \begin{equation}
    \label{eq:invcov}
    \textit{\~C}^{-1}_{ij}=\frac{N_{mock}-N_p-2}{N_{mock}-1}C^{-1}_{ij}
  \end{equation}
where, $N_p$ represents the number of data points used. The Bayes factor between the BigMDPL-QSO and the BigMDPL-QSOZ model is
\begin{equation}
    \label{eq:bayes}
    K=\frac{P(\xi_{data}|\xi_{\textsc{qsoz}})}{P(\xi_{data}|\xi_{\textsc{qso}})} = 5.45.
  \end{equation}

This result suggests that BigMDPL-QSOZ model is more substantially supported by the data than BigMDPL-QSO. The Bayes factor between the BigMDPL-QSOZ and the BigMDPL-QSO-NSAT is $K=1.67$. In this case, we cannot conclude which model better reproduces the data. Furthermore the BigMDPL light-cones have an important variability between realisations when the random seed is changed and it is not possible to construct a sufficient number of independent light-cones to make a definitive statement about the two models. In terms of $\chi^2$ both light-cones are in agreement with the current data, though including a model with more parameters will improve the fitting of the data.

\subsection{Checking $\xi_2(s)$ and $w_p(r_p)$}
The quadrupole is very sensitive to processes affecting the small scales. Effects due to fiber collisions have an important impact at scales beyond the fiber size. However, the effect of fiber collisions is very small in the QSO sample. The most important observational effect is due to redshift errors, as shown in Appendix \ref{app:resolution}. Figure \ref{fig:quad} shows the quadrupole of the BigMDPL-QSO, BigMDPL-QSOZ and BigMDPL-QSO-NSAT light-cones compared to the observations. All light-cones reproduce the data within 1-$\sigma$ error. This agreement suggests that we are using a reasonable model to account for redshift errors. We note that the BigMDPL-QSOZ light-cone reproduces the quadrupole better than the other two light-cones.
\begin{figure}
 \includegraphics[width=84mm]{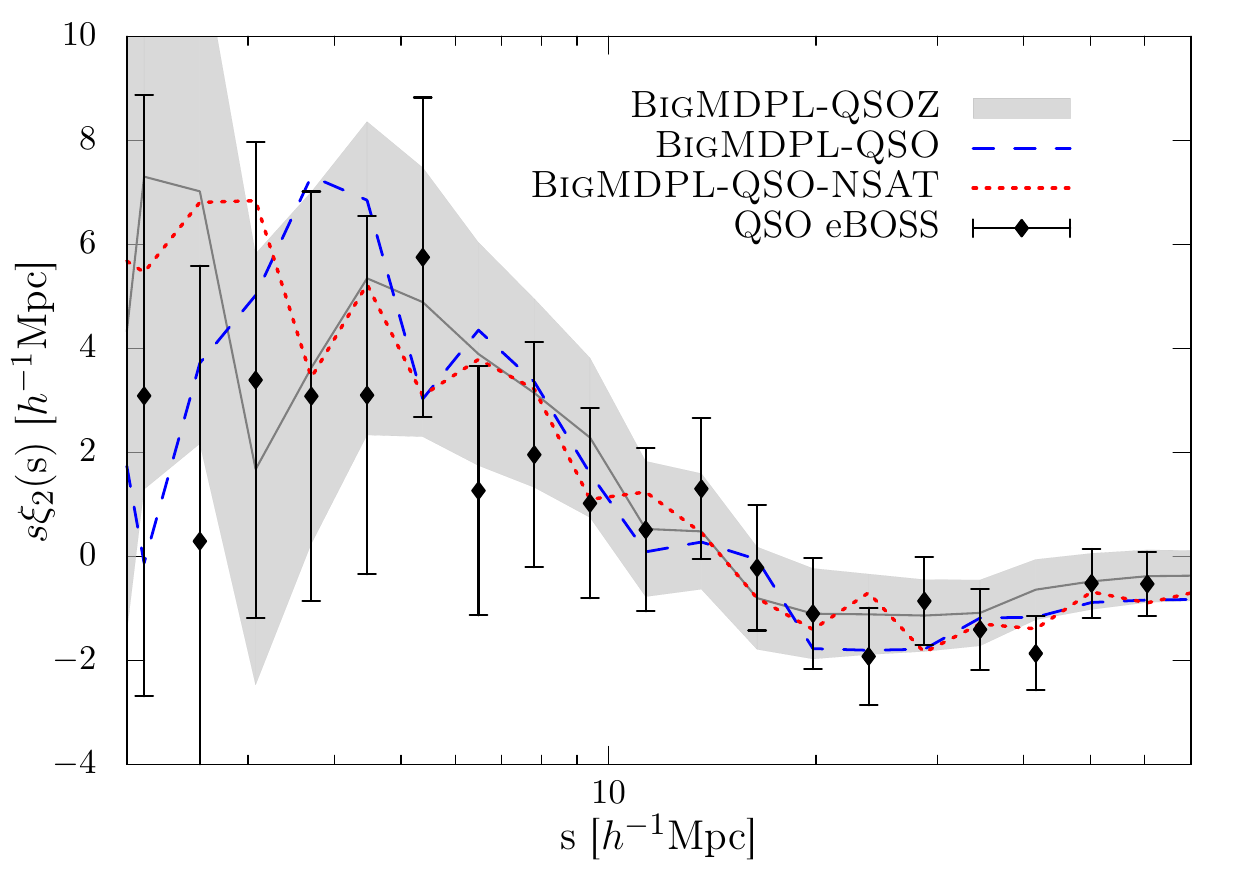}
 \caption{Quadrupole vs. comoving scale in redshift-space predicted by the BigMDPL-QSOZ (shaded region), BigMDPL-QSO (dashed line) and BigMDPL-QSO-NSAT (dotted lines) compared to the Y1Q (black points). All mocks are in agreement with observations. Error bars and shaded areas are computed using 1000 GLAM catalogues and correspond to $1-\sigma$ deviation from the mean value.}
 \label{fig:quad}
\end{figure}

\begin{figure}
 \includegraphics[width=84mm]{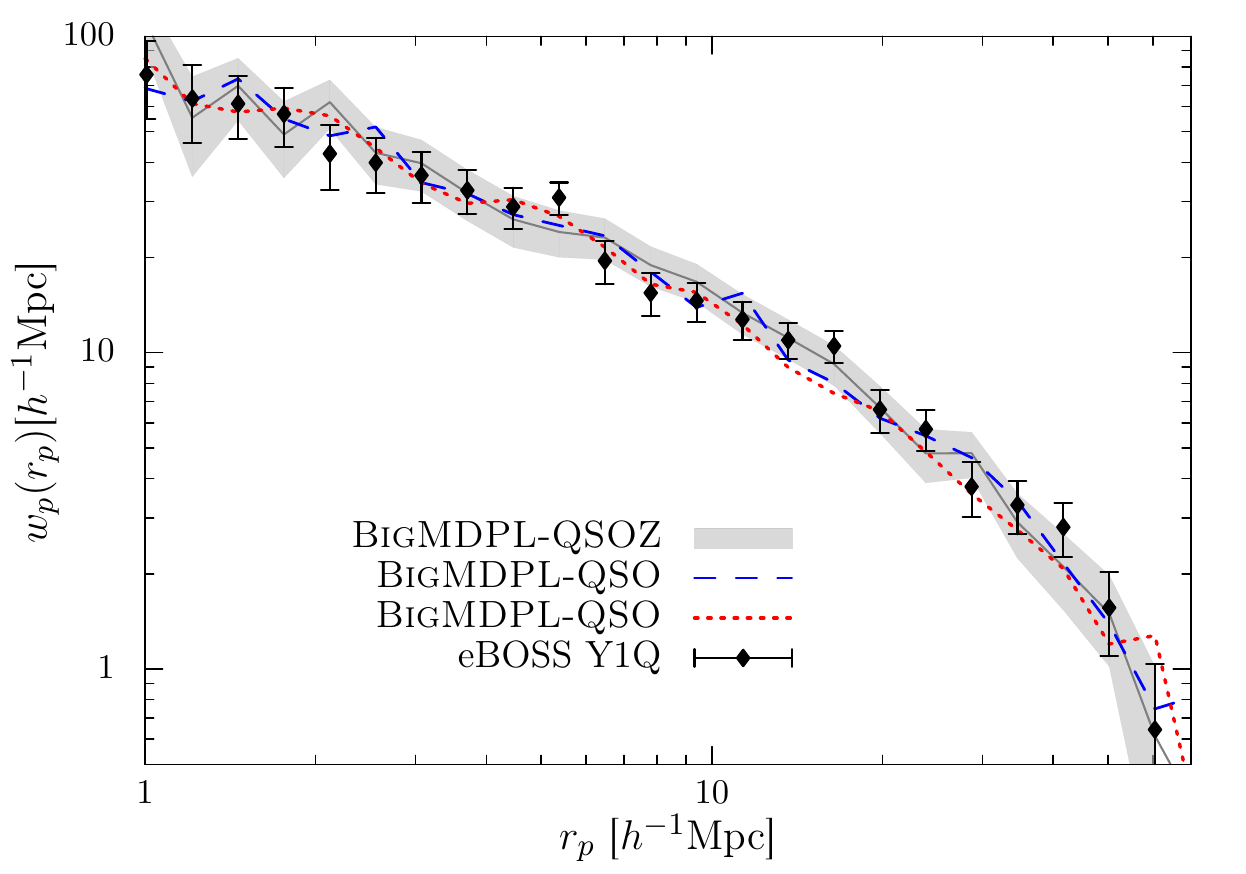}
 \caption{Projected correlation function predicted by the BigMDPL-QSOZ (shaded region), BigMDPL-QSO (dashed line) and BigMDPL-QSO-NSAT (dotted line) compared to the Y1Q (black points). The width of the shaded area represents 1-$\sigma$ errors computed with 1000 GLAM catalogues and correspond to $1-\sigma$ deviation from the mean value. Our model reproduces the clustering for all relevant scales.}
 \label{fig:wp}
\end{figure}

We compared the projected correlation function for the three light-cones and the observed data, finding a good agreement shown in Figure\ref{fig:wp}. 

The clustering predicted by the best-fit model, which is mainly determined by the $V_{mean}$, reproduces with good agreement the 2-point statistics of the observed data. We do not find significant differences between the three light-cones presented, all of them can reproduce the 2-point statistics of the complete Y1Q sample with good agreement.

\subsection{Bias}

The Y1Q data allows for accurate measurements of the correlation function $\xi(r)$ and of the quasar bias $b_{Q}$, within the redshift range $0.9 < z < 2.2$. Laurent et al. in prep. obtain $b_{Q} = 2.45\pm0.05$, when averaged over separations between 10 and 90 $h^{-1}$Mpc. This value is compatible with previous SDSS measurements,  $b_Q(z=1.58)=2.42\pm0.40$, by \citet{ross2009}.

\begin{figure}
 \includegraphics[width=84mm]{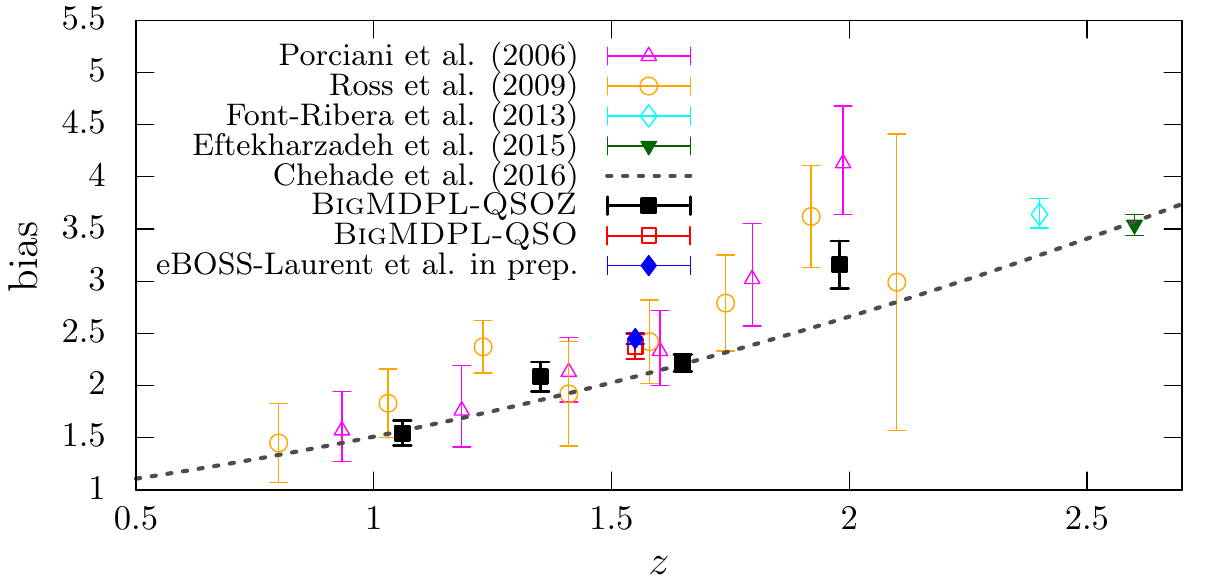}
 \caption{QSO bias as a function of redshift. The bias is computed using BigMDPL-QSOZ and BigMDPL-QSO light-cones. We include results from \citet{Chehade2016}, \citet{eftekharzadeh2015},\citet{font-ribera2014}, \citet{ross2009} and \citet{porciani2006}. eBOSS bias measurements are in agreement with previous results and about ten times more precise. Results of eBOSS from Laurent et al. in prep. are also included.}
 \label{fig:bias}
\end{figure}

We estimate the bias using the dark matter counter-part of the QSO mock light-cone. Using the autocorrelation of the dark matter sample, and the correlation function of the QSO mock in real space, we estimate the bias using 
\begin{equation}
\label{eq:bias}
b(r)^2=\frac{\xi(r)}{\xi_{\textsc{DM}}(r)}.
\end{equation}
Figure \ref{fig:bias} presents the bias of the BigMDPL-QSOZ and the BigMDPL-QSO compared to previous studies. 

The bias measurements presented in Figure \ref{fig:bias} come from spectroscopically confirmed quasars in the two degree field \citep[2dF;][]{porciani2006} at $0.8<z<2.1$, SDSS-I/II \citep{ross2009} at $z<2.2$, the Quasar Dark Energy Survey pilot \citep[2QDESp;][]{Chehade2016} for redshift between 0.8 and 2.5 and the BOSS sample \citep{eftekharzadeh2015} at $2.2<z<2.8$. All these studies parametrise the real space correlation function by a power-law, $\xi(r)=(r/r_0)^\gamma$, which can be related with the observed correlation function (redshift space) by
  \begin{equation}
  \xi(s)=\bigg(b_\textsc{q}^2+\frac{2}{3}b_\textsc{q}f+\frac{f^2}{5}\bigg)\xi(r),    
  \end{equation}
where $f=[\Omega_m(z)]^{0.56}$ is the gravitational growth factor. In addition, we include measurements of quasars via Lyman-$\alpha$ absorption at redshift 2.4 from the BOSS sample \citep{font-ribera2014}. \citet{eftekharzadeh2015} also show a comparison between different estimations of the bias. At the redshifts studied, the bias measurements obtained in our study are in good agreement (see Figure \ref{fig:bias}) and they are a factor 5 to 10 times more precise than previous studies.

\subsection{Cross-correlation coefficients}

\begin{figure}
 \includegraphics[width=84mm]{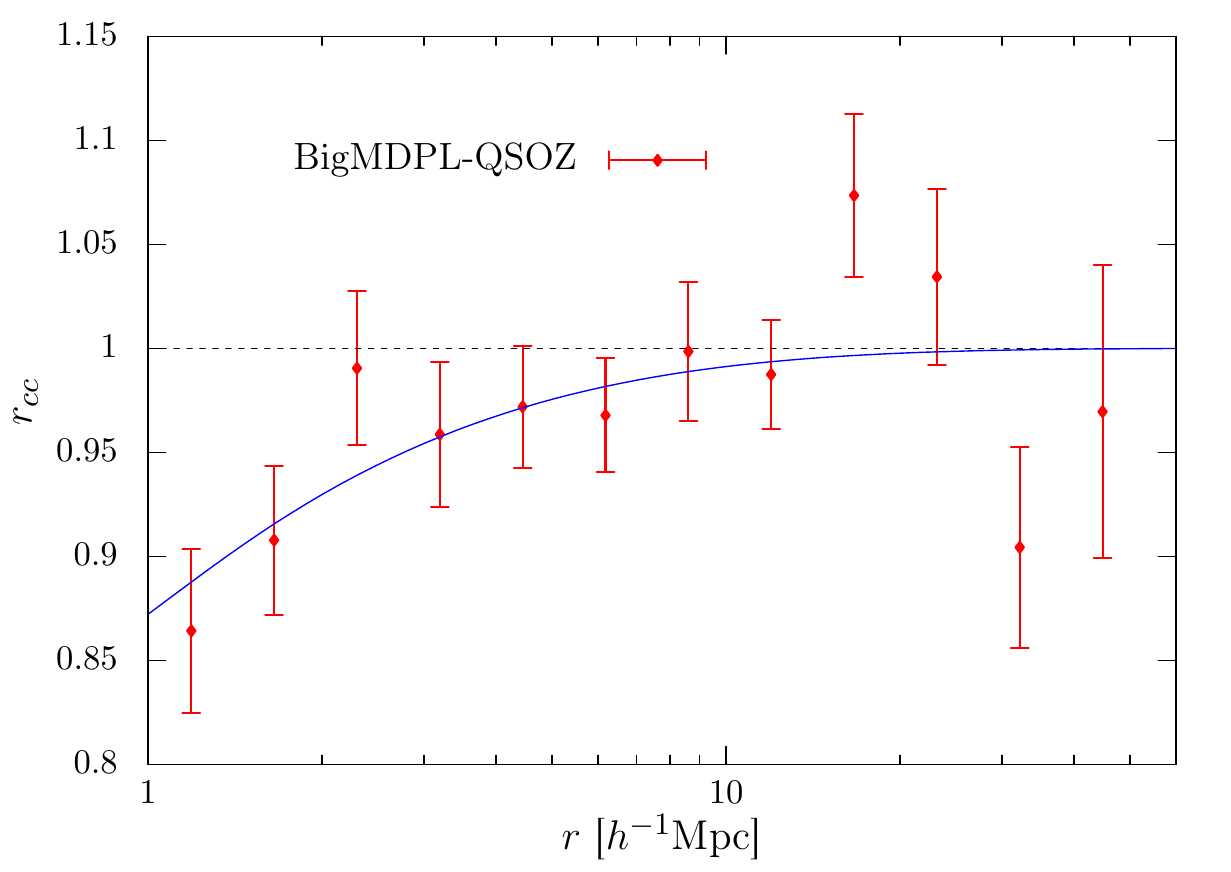}
 \caption{Cross-correlation coefficient between the dark matter field and the BigMDPL-QSOZ light-cone. The best model from \eqref{eq:perteo} is shown with a solid line.}
 \label{fig:coeff}
\end{figure}

The linear bias provides a good description of the relationship between dark matter and QSO mock in the linear regime. However, a single parameter $b_Q$ is not enough to understand the link between galaxies and dark matter at all scales. To parametrise this relationship, we use the second order bias, which is related to scales smaller than 10 $h^{-1}$ Mpc. The second order bias is inferred from the cross-correlation coefficient. It gives an estimation of the correlation between the positions of quasars and the dark matter field \citep{dekel1999}. The cross-correlation, denoted $r_{cc}$, between quasars and the dark matter field is defined as
\begin{equation}	
r_{cc}(r)=\frac{\xi_{qm}(r)}{\sqrt{\xi_{qq}(r)\xi_{mm}(r)}},
\end{equation}
where $q$ denotes the quasar sample and $m$ the dark matter. $r_{cc}$ is sensitive to the non-linear stochastic bias of the sample. Figure \ref{fig:coeff} shows the cross-correlation coefficient between BigMDPL-QSOZ and the dark matter field. For scales larger than 10 $h^{-1}$ Mpc, the cross-correlation function is consistent with 1. As expected, in this regime, we have  $\xi_{gm}=b_Q\xi_{mm}$ and $\xi_{gg}=b_Q^2\xi_{mm}$. 
At smaller separations, $r_{cc}$ becomes smaller than one. This tendency is described in perturbation theory \citet{baldauf2010}, where $r_{cc}$ is described with the second order bias by 
\begin{equation}
\label{eq:perteo}
r_{cc}(r)\approx1-b_2^2\frac{\xi_{lin}(r)}{4},
\end{equation}
where $b_2$ is the second order bias and $\xi_{lin}$ is the linear correlation function. The cross-correlation coefficient fit directly to the clustering by $b_2=0.314\pm0.030$. This relation is sufficient for the scales studied ($1<rh^{-1}$ Mpc $<10$), see the solid line in Fig. \ref{fig:coeff}.

\subsection{Halo Occupation Distribution}

Table \ref{table:hod} shows the mean mass of halos hosting quasars, the satellite fraction characterises how quasars populate dark matter halos and the mean value of $V_{max}$ for all light-cones built in this study. 
\begin{table}
\centering
\caption{Mean halo mass and satellite fraction prediction from the BigMDPL light-cones.}
\label{table:hod}
\begin{tabular}{cccc} \hline \hline
Light-cone& $V_{mean}$ & $\log_{10}[M_{200}/M_\odot]$ & $f_{sat}$\\ 
 & [s$^{-1}$Km] & & \\ \hline 
 BigMDPL-QSOZ  & 326.9 & 12.61 & 0.048\\
 BigMDPL-QSO & 341.2 & 12.66 & 0.053\\ \
 BigMDPL-QSO-NSAT & 349.5 & 12.70 & 0.0\\ \hline 
\end{tabular}
\end{table}

If the satellite fraction is not fixed (no distinction between halos and sub halos), we obtain a non-negligible fraction of satellites, $\sim5$\%. This value is consistent with \citet{shun2013} which find a satellite fraction of 6.8\%. However, due to the degeneracy between $V_{mean}$ and $f_{sat}$, our model could also match the clustering with a negligible fraction (Figure \ref{fig:chi2}), as presented in \citet{richardson2012}.

\begin{figure}
 \includegraphics[width=84mm]{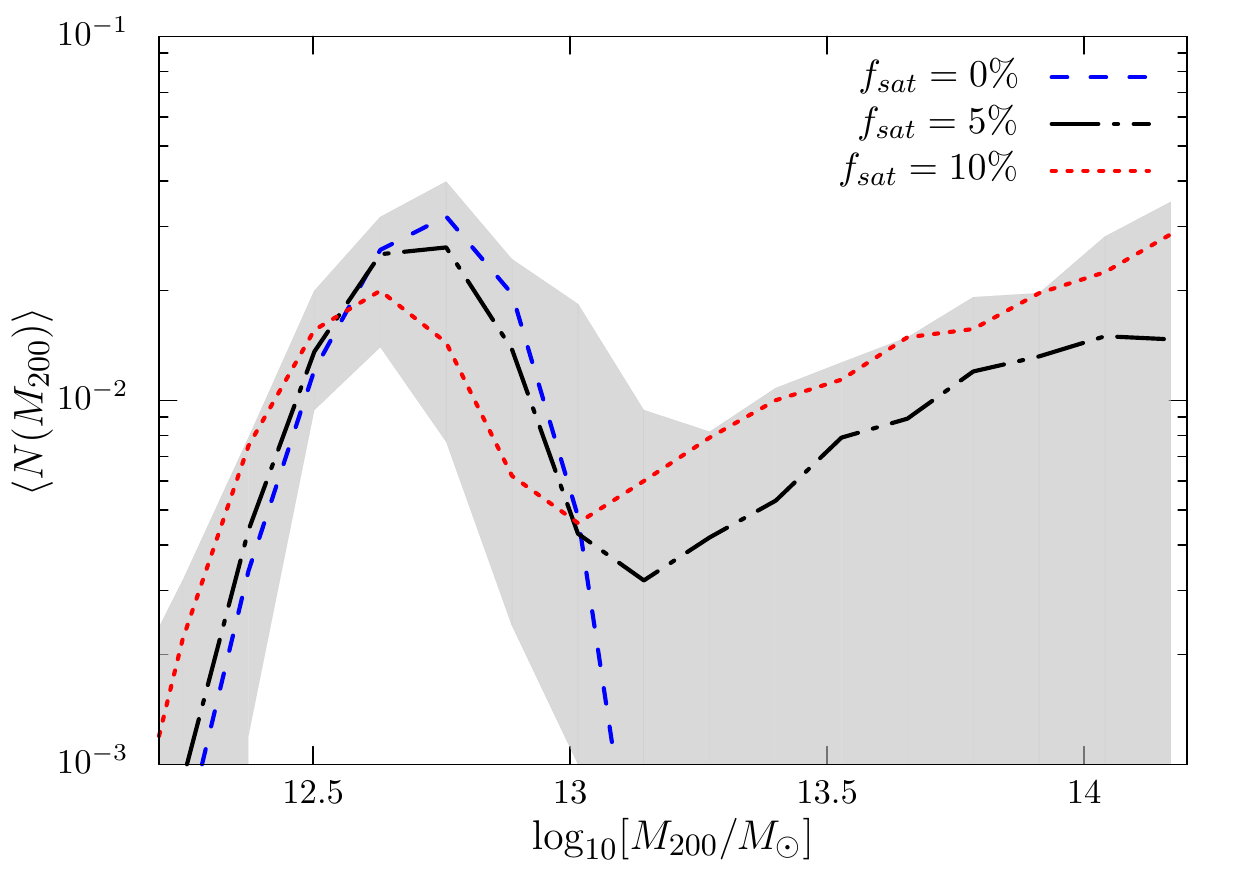}
 \caption{Halo Occupation distribution for central plus satellites predicted from the BigMDPL-QSO light-cone. We present three light-cones using different fraction of satellites. The shaded area is computed adding 1-$\sigma$ error in the $V_{mean}$ parameter for each light-cone. In addition, we vary the width of the distribution from 10 to 60 s$^{-1}$km to see the impact of this parameter in the HOD. $f_{sat}$ is also changed from 0 to 0.12}
 \label{fig:hod}
\end{figure}

Another way to formulate how QSO populate the density field is the probability of finding $N$ quasars in a halo of mass $M$ ($\big<N(M)\big>$), namely the HOD model. This method describes how quasars would statistically populate halos using a set of parameters fitted directly on the clustering. In SHAM models, $\big<N(M)\big>$ is given by the halo catalogue by counting the total number of host halos and the number of QSO per bin of mass. Figure \ref{fig:hod} shows the halo occupation distribution predicted by the BigMDPL-QSO light-cone. We use this light-cone rather than the other as it has a negligible fraction of objects from the incomplete part of the BigMDPL simulation. It also allows $\sigma_{max}$ and $f_{sat}$ to vary in a wide range, letting us show the dependency of $\big<N(M)\big>$ on these parameters reflected in the different lines of Figure \ref{fig:hod}.

Additionally, we construct light-cones with different $V_{mean}$ including variations of 1-$\sigma$ from the best-fit. We also vary the width of the distribution between 10 and 60 s$^{-1}$km. We do not use a larger $\sigma_{peak}$, because we do not want to include a large fraction of objects coming from the incomplete part of the simulation. $f_{sat}$ also varies between 0 and 10\%. The shaded area in Figure \ref{fig:hod} represents all HODs encompassed by these parameter variations.

Compared to previous HOD results \citep{shun2013}, our model puts new constraints for masses below $10^{13}$ $M_\odot$. We find a distribution dominated by the mean halo mass of the sample. However, $\big<N(M)\big>$ has a strong dependency with the other two parameters of the model, which we cannot constrain with the current data. An improvement on small scales of the QSO clustering or the cross correlation between ELG and QSO in future surveys would constrain $\sigma_{peak}$ and $f_{sat}$ and therefore provide better HOD predictions.

\section{Discussion}
\label{sec:disc}

Previous HOD analysis of the SDSS QSO sample combined different data sets to get more information about the distribution of QSOs inside halos. However, due to large uncertainties in the data, the parameters of the HOD remain degenerate. eBOSS will greatly increase the statistical size of quasar samples, giving an excellent opportunity to learn more about this population and its connection with the dark matter. What we do here is to present the first study of the Y1Q clustering introducing a modified halo abundance matching which allows us to predict the HOD, masses of the dark matter halos and the bias of the sample.

Several studies have provided information about quasars at different redshifts using their clustering measurements. \citet{richardson2012} study the clustering of the 48,000 QSO from the SDSS sample in the redshift range $0.4<z<2.5$. They interpret the measurements of the projected correlation function at redshift 1.4. In addition, 4,426 spectroscopically identified quasars in the redshift interval $2.9< z <5.4$ \citep{shen2007} are used to study the small scale clustering. However, they use a regular HOD without including a duty cycle. For this reason, their parameters reproduce the clustering, but most of them are unphysical.  \citet{shun2013} study the two-point cross-correlation function of 8,198 SDSS QSO and 349,608 BOSS CMASS galaxies in the redshift range $0.3 < z < 0.9$. They provide predictions of the HOD from quasars. However, the large degeneracies of the parameters make it impossible to have a well constrained HOD. The BOSS sample provides a set of CORE QSO which are studied by \citep{eftekharzadeh2015}. They extend the analysis of the projected correlation function of the BOSS sample done by \citet{white2012}. In that analysis, $\sim70,000$ quasars in the redshift range 2.2 to 3.4 are studied. In a more recent study, \citet{Chehade2016} combine the optical photometry of the 2dF Quasar Dark Energy Survey pilot (2QDESp) and the bands of the Wide-field Infrared Survey Explorer (WISE) to provide a sample of $\sim$10,000 QSO in the redshift range 0.8 to 2.5. Our study uses a larger and wider QSO sample than in previous works. It allows us to have a good estimation of the clustering in the redshift range $0.9<z<2.2$.

The mean mass of halos hosting quasars has been measured by different methods finding a reasonable agreement between their results. However, the range of masses cover by quasars is still not well constrained. \citet{richardson2012} predict a mean halo mass for central halos $M_{cen}\sim 10^{12.77} M_\odot$ with a small fraction of QSO satellites, $7.4\times 10^{−4}$. This result is in agreement with the BigMDPL-QSO-NSAT, which provides host halo masses for quasars of  $10^{12.7\pm0.16}$ M$_\odot$. \citet{shun2013} model the cross-correlation between CMASS galaxies and QSO by a power-law, $\xi_{QG} = (r/r_0)^\gamma$, with $r_0=6.61\pm 0.25$ $h^{-1}$ Mpc and $\gamma = 1.69 \pm 0.07$ for scales $r=2-25$ $h^{-1}$ Mpc. They find a characteristic mean halo mass of $10^{12.8}$ $M_\odot$. In contrast to \citet{richardson2012}, a non-negligible satellite fraction is predicted by \citet{shun2013}. They find that $6.8$ per cent of QSO are hosted by sub-halos. This result is in better agreement with our mocks without fixing the fraction of satellites, which predict $\sim$5\% of quasars living in sub-halos. The halo masses predicted by this HOD are also in agreement within 1-$\sigma$ errors with our measurements. Nevertheless, they have larger degeneracies between their parameters. From the BOSS sample, \citet{white2012} find the quasar halo masses covering a wide mass range between $10^{11.59}$ $M_{\odot}$ and $10^{12.65}$ $M_{\odot}$. Just as in the previous cases, these values of masses are still in agreement with our results shown in Table \ref{table:meanVpeak}. \citet{Chehade2016} results are compared with other surveys (SDSS, 2QZ and 2SLAQ). As in previous works, they find no evidence of a dependency between the clustering and the luminosity of the QSO. In addition, they show that quasar clustering depends on redshift, in particular, when BOSS data is included. They describe the clustering of the sample using a power-law, where $r_0=7.3\pm0.1$ $h^{-1}$ Mpc at redshift 2.4, while the correlation scale for the whole redshift range is $r_0=6.1\pm0.1$ $h^{-1}$. Their measurements are consistent with host halos masses of $\sim 10^{12.46}$. Future observations will allow cross-correlations between ELGs and quasars, which will enable a better understanding of the distribution of quasars within the dark matter halo. These measurements could fix the satellite fraction of quasars. However, the width of the distribution is more difficult to constrain. In the similar case of ELG, \citet{favole2015b} faced an equivalent problem to describe their clustering. They use constraints from lensing measurements to understand the clustering on the smallest scales. Unfortunately, such measurements are not available for quasars. 
\begin{table}
\centering 
\caption{Mass prediction of halos hosting quasars for different samples. It is presented with the name of the method used to analyse the sample and the used redshift range. $^1$This work, $^2$\citet{shun2013}, $^3$\citet{richardson2012}, $^4$\citet{white2012}, $^5$\citet{eftekharzadeh2015}, $^6$\citet{Chehade2016}}
\label{table:hod}
\begin{tabular}{ccccc} \hline \hline 
  Sample & $N_\textsc{qso}$ & $z$ & Method & $\log_{10 }(M_{h}/M_\odot)$ \\ \hline  
  eBOSS$^1$ & 68,269 & 0.9--2.2 & HAM & 12.5--12.82\\
  SDSS-I/II$^2$ & 8198 & 0.3--0.9 & Power-law fit & 12.75\\ 
  SDSS-I/II$^3$ &  48,000 & 0.4--2.5 & HOD & 12.70--12.77\\ 
  BOSS$^4$ & 27,129 & 2.2--2.8 & Power-law fit &  12.59--11.65\\ 
  BOSS$^5$ & 55,826 & 2.2--2.8 & Power-law fit & 11.63-12.63 \\
  2QDESp$^6$ & 10,000 & 0.8--2.5 & Powe-law fit & 12.17--12.64 \\ \hline 
\end{tabular}
\end{table}

Using our model, the signal of the clustering in the linear regime is dominated by the mean halo mass of the distribution. This is clear in the halo occupation distribution (Figure \ref{fig:hod}), where the distribution has a strong peak near the mean halo mass of the sample. We find a more constrained HOD region for quasars than \citet{shun2013}. However, more information from small scales is needed to have better constraints in the satellite fraction and width of the distribution in order to provide more realistic uncertainties. We find a bias equal to $2.37\pm0.12$ for the redshift range $0.9<z<2.2$, which is in good agreement with previous analysis and with eBOSS data from Laurent et al. (in prep) (Figure \ref{fig:bias}). We provide measurements for the evolution of the bias using the BigMDPL-QSOZ light-cone, finding that the eBOSS quasars are in agreement with $b_Q$=1.54, 2.08, 2.21, 3.15 for redshift 1.06, 1.35, 1.65, 1.98. Furthermore, to give a complete parametrisation of the scales studied in this work, we calculate the second order bias from the cross-correlation coefficients, finding $b_2=0.314\pm0.030$. Table \ref{table:hod} presents a comparison of the halo mass predictions of previous studies and our result.

\section{Summary}
\label{sec:summary}
We modelled the clustering of $\sim$70,000 optical quasars from the eBOSS Y1Q \textsc{core} sample in the redshift range $0.9<z<2.2$. We used a modified halo abundance matching that takes into account the incompleteness of the QSO sample and the intrinsic scatter between QSOs and dark matter halos. This model was implemented in a light-cone constructed from a 2.5 $h^{-1}$Gpc simulation, covering an area comparable to the eBOSS Y1Q sample. 

Our main results can be summarised as follows.
\begin{itemize}
\item We assume that the $V_{max}$ distribution of halos hosting QSOs is described by a Gaussian function which is defined by its mean and width plus one parameter for the satellite fraction. The current observations do not bear information on small scale clustering. For this reason, we cannot constrain the fraction of satellites. Hence, we do not distinguish between host and sub halos when the selection is done. The final mock thus has the same fraction of satellites as the complete simulation in the mass range used. 

\item We model the clustering of the Y1Q using a single free parameter ($V_{mean}$). The width of the Gaussian distribution is fixed to 30 s$^{-1}$ km and we only impose a value to the satellite fraction in the BigMDPL-QSO-NSAT light-cone, for the others light-cones we do not fix this parameter.

\item The prediction of our model is in a good agreement with the 2PCF and the monopole of the power spectrum of the Y1Q data. The light-cone is constructed assuming Gaussian redshift errors given by \citet{dawson2016}. Their modelling improves the agreement between our model and the data. It provides a good description of the observed clustering on small scales, which is very sensitive to variations caused by these errors.

\item We construct three kinds of light-cones: one including the evolution of the parameters with redshift (BigMDPL-QSOZ), another describing the whole redshift range with a single parameters (BigMDPL-QSO) and a third one fixing the satellite fraction to zero (BigMDPL-QSO-NSAT). The mean halo masses are $10^{12.61}$ $M_\odot$, $10^{12.66}$ $M_\odot$ and $10^{12.70}$ $M_\odot$  respectively.

\item Using the Bayes factor we find a strong evidence that the BigMDPL-QSOZ (4-parameters) reproduces the data better than the BigMDPL-QSO (1-parameter). However, we cannot make the same conclusion with the model without satellites, which reproduces the data with a similar agreement to the BigMDPL-QSOZ model.

\item We find a mean bias of the Y1Q sample equal to $2.37\pm0.12$ and a second order bias $b_2=0.314\pm0.030$, which both describe the relation between the dark matter and the QSO mock for the studied scales. 
\end{itemize}

BigMDPL-QSOs and GLAM-QSO eBOSS mocks are publicly available through the $\textit{Skies and Universes}$ website \footnote{\url{http://projects.ift.uam-csic.es/skies-universes}}.

%%%%%%%%%%%%%%%%% REFERENCES %%%%%%%%%%%%%%%%%%

\bibliographystyle{mnras}
\bibliography{references.bib}

%%%%%%%%%%%%%%%%%%%%%%%%%%%%%%%%%%%%%%%%%%%%%%%%%%

% Don't change these lines
\bsp	% typesetting comment
\appendix

\section{Simulation resolution}
\label{app:resolution}

In order to reproduce the observed clustering of QSO or ELG samples, simulations with large volume and a high resolution are needed to resolve halos of masses $\sim10^{12.5}M_\odot$. The Y1Q sample covers $\sim1100$ deg$^2$ of the sky. This area is comparable to the BigMDPL-QSO light-cone. However, a small part of the halo mass range occupied by quasars can be in the incomplete part of the simulation. 

We use the 1 $h^{-1}$Gpc MDPL simulation to quantify the effect of incompleteness of the BigMDPL light-cone. We select two snapshots from each simulation with similar redshift (Table \ref{table:resolution}). We apply the model using the parameters of Table \ref{table:meanVpeak}. 
Table \ref{table:resolution} presents a comparison between both simulations. In terms of halo mass, mocks constructed with both simulations provide consistent mean halo masses. Similar results are found for the satellite fraction.

\begin{figure}
 \includegraphics[width=84mm]{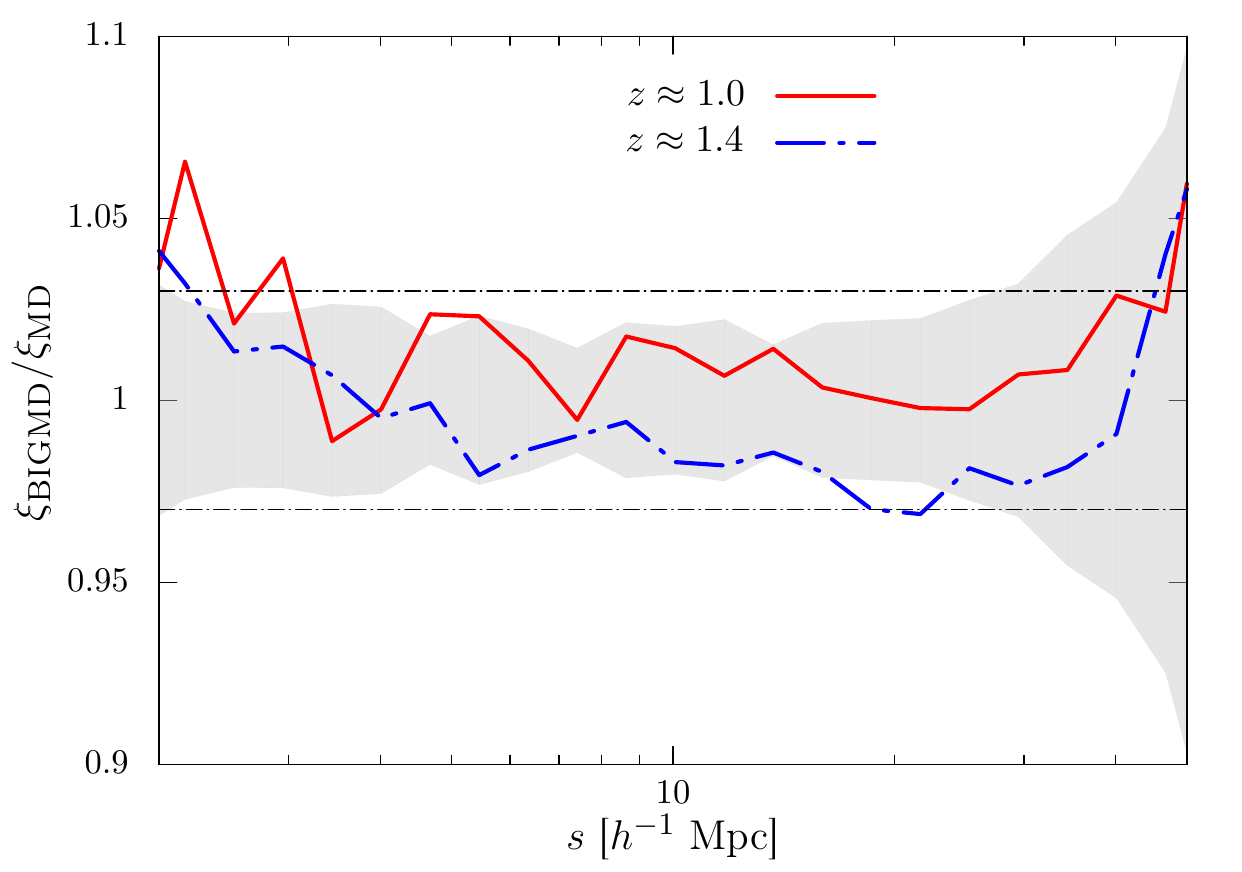}
 \caption{Ratio between BigMDPL and MDPL mocks of the monopole of the correlation function in configuration space. The horizontal lines represent 3\% differences. The shaded area shows 1-$\sigma$ dispersion due to the random selection in the MDPL boxes. We use 15 realisations to compute the shaded area.}
 \label{fig:resolution}
\end{figure}

In terms of clustering, both simulations give coherent results with differences of the order of 3\%. Figure \ref{fig:resolution} shows the difference on the monopole between both simulations. These discrepancies are not a problem for our analysis, where errors from the data are of the order of 15\%.
\begin{table}
\centering
\caption{Comparison of the halo mass of mocks constructed with the BigMDPL and MDPL simulations. For comparison, all snapshots of the BigMDPL simulation in the redshift range $0.9<z<2.2$ were used. We select snapshots with the nearest redshift from the MDPL simulation}
\label{table:resolution}
\begin{tabular}{ccccc} \hline 
 Box & z & $\log_{10}[M/M_\odot]$ & $V_{mean}$ & $f_{sat}$\\ \hline
 \multirow{2}{1cm}{MDPL} & 0.987 & 12.41 & 284.25 & 0.08\\
& 1.425 & 12.54 & 325.95 & 0.07\\ \hline
\multirow{2}{1cm}{BigMDPL} & 1.000 & 12.40 & 284.25 & 0.11\\
& 1.445 & 12.55 & 325.95 & 0.07\\ \hline
\end{tabular}
\end{table}

In addition to the large errors in the data, discrepancies between both boxes seem reasonable if we notice the other sources of error.
\begin{enumerate}
\item Both simulations have different initial conditions, this includes variations due to the cosmic variance between simulations.
\item The shot noise in the correlation function is larger in the MDPL simulation due to the smaller volume.
\item The random selection of our model is another source of errors. The shaded area in Figure \ref{fig:resolution} represents the 1-$\sigma$ dispersion of 15 mocks produced with different seeds.
\item The BigMDPL simulation includes long-waves which are not included in the 1 $h^{-1}$Gpc box size.
\end{enumerate}

\section{Effects of observational errors on the clustering}
\label{app:errors}

\begin{figure}
 \includegraphics[width=84mm]{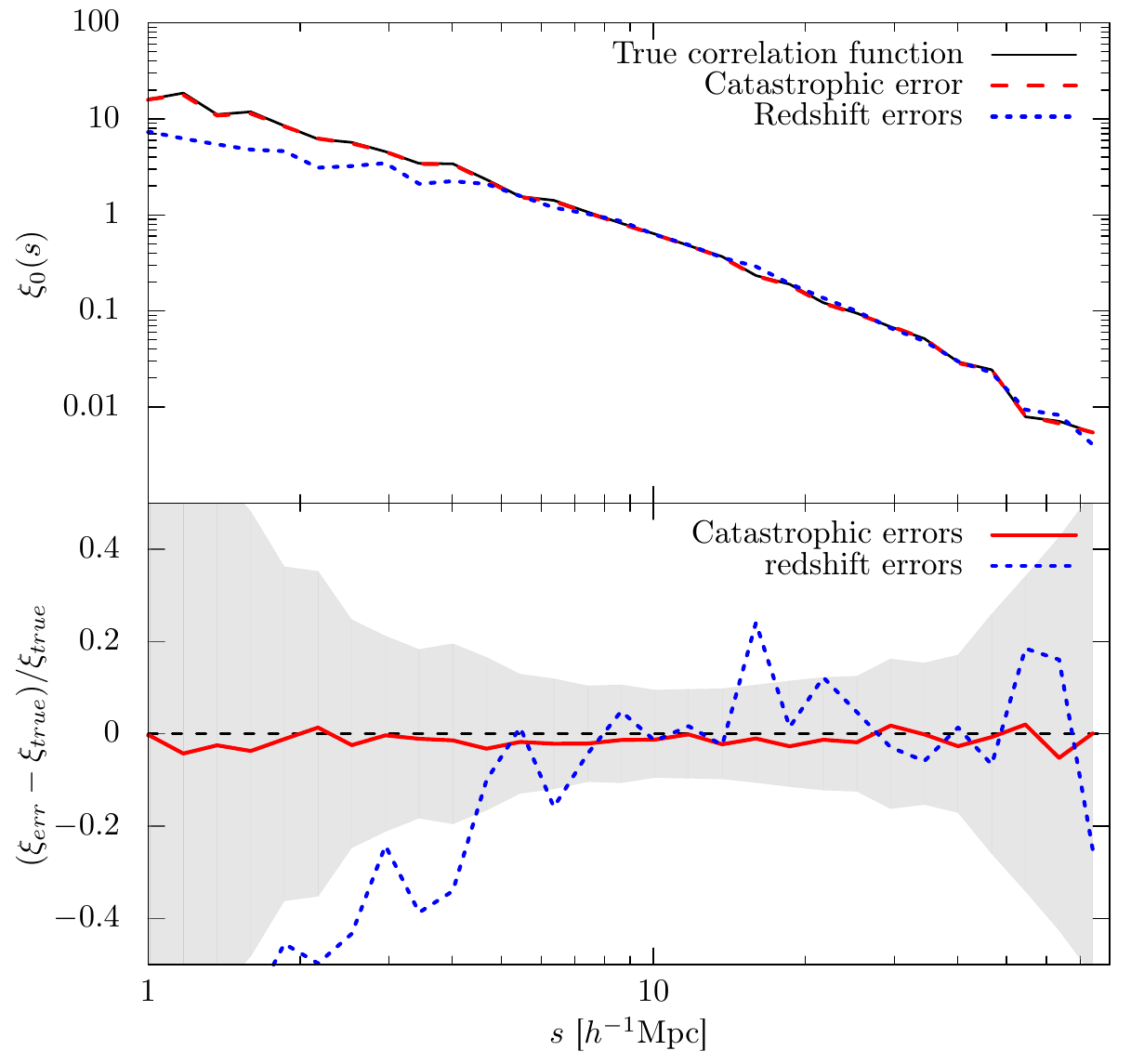}
 \caption{\textit{Top panel:} Impact of catastrophic redshift errors and redshift errors on the monopole of the correlation function. A light-cone reproducing the Y1Q 1-point and 2-point statistics is used for this comparison. \textit{Bottom panel:} Normalised differences between mocks including redshift errors (blue dotted line) and catastrophic redshift errors (red line) with a model without errors. The shaded area represents the statistical errors in the light-cone computed from 1000 GLAM catalogues. Differences due to catastrophic redshift errors are $\sim$1\%. Redshift error have an important impact at small scales which cannot be explain by uncertainties from mocks.}
 \label{fig:rdserr}
\end{figure}

The model presented in this work includes two observational errors: Catastrophic redshift errors and redshift errors. The first errors cause a constant reduction in the clustering amplitud at all the scales. Figure \ref{fig:rdserr} shows the effect of applying 1\% of catastrophic redshifts. We find a reduction of $\sim$1\% in all scales of the correlation function in configuration space. 

Redshift errors have the strongest impact on the clustering. The selection of QSO implies fixing maximum width (precision) to identify the emission/absorption features of the spectra. We introduce the effect of this tolerance using Gaussian errors with a width given by \citet{dawson2016}. Redshift errors have an important impact at scales $<10h^{-1}$Mpc. In Figure \ref{fig:rdserr}, it is possible to see a disagreement larger than 40\%, which cannot be explained by statistical errors of the sample (shaded area Figure \ref{fig:rdserr}).
\begin{figure}
 \includegraphics[width=84mm]{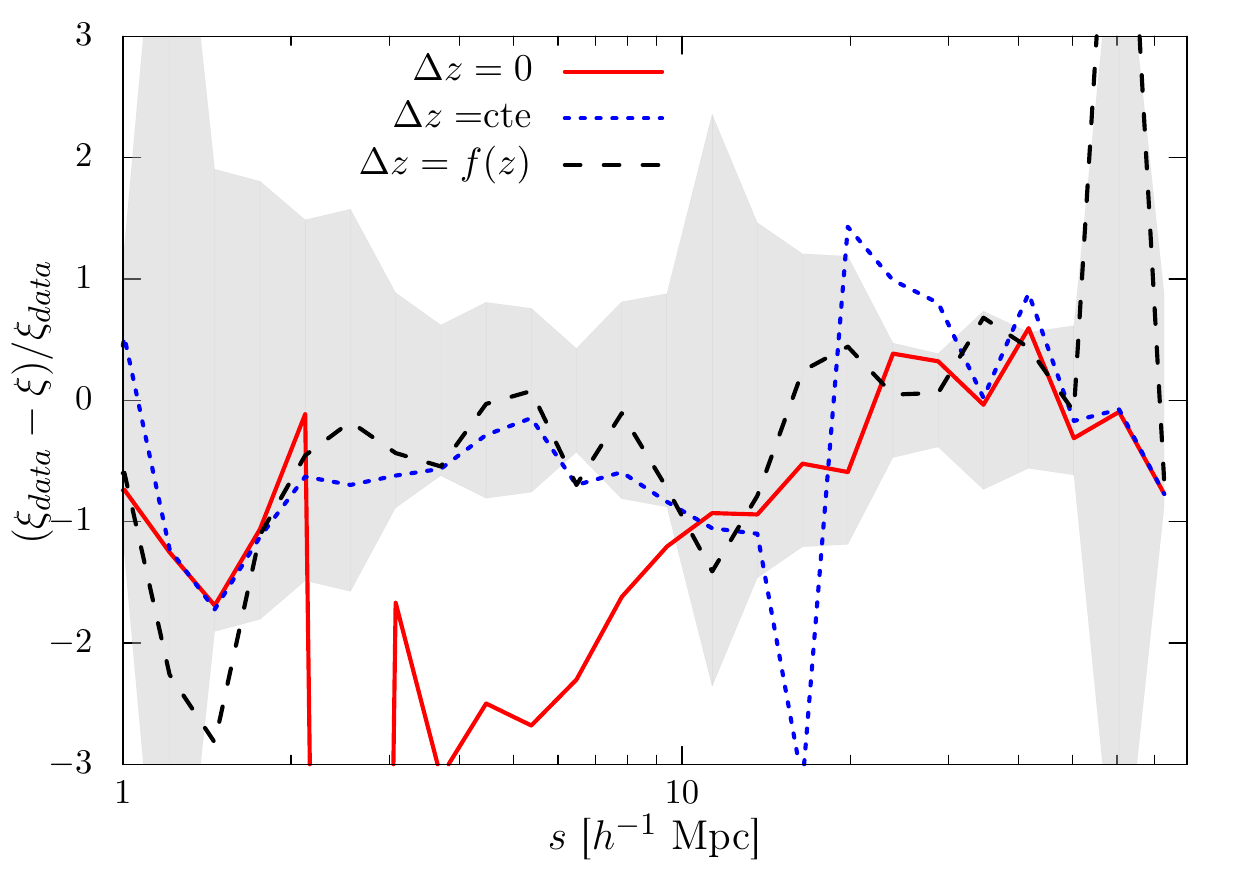}
 \caption{Impact of redshift errors in the quadrupole of the correlation function in configuration space. Lines show the normalised difference between observed data and model without redshift errors (red solid line), constant redshift error $\Delta z=0.005$ (blue dotted line) and including redshift errors given by equation \eqref{eq:rdserr} (black dashed line). Shaded area represent 1-$\sigma$ error computed with 1000 GLAM catalogues. for one light-cone.}
 \label{fig:quadrdserr}
\end{figure}

The impact of redshift error is very important in the monopole of the correlation function. However, the effects on the quadrupole are larger. Figure \ref{fig:quadrdserr} shows the ratio of quadrupole from the observed data and the different mocks. The model introduced in this work describes the very large difference found between our mock and the observed data.

\section*{Acknowledgements}
SRT is grateful for support from the Campus de Excelencia Internacional UAM/CSIC. 

SRT, JC, FP acknowledge support from the Spanish MICINN Consolider-Ingenio 2010 Programme under grant MultiDark CSD2009-00064 MINECO Severo Ochoa Award SEV-2012-0249 and grant AYA2014-60641-C2-1-P. 

GY acknowledges  financial support from MINECO/FEDER  (Spain) under research grants AYA2012-31101 and AYA2015-63810-P.

The \textsc{BigMultiDark} simulations have been performed on the SuperMUC supercomputer at the Leibniz-Rechenzentrum (LRZ) in Munich, using the computing resources awarded to the PRACE project number 2012060963. The authors want to thank V. Springel for providing us with the optimised version of GADGET-2.

Funding for the Sloan Digital Sky Survey IV has been provided by the Alfred P. Sloan Foundation, the U.S. Department of Energy Office of Science, and the Participating Institutions. SDSS acknowledges support and resources from the Center for High-Performance Computing at the University of Utah. The SDSS web site is www.sdss.org.

SDSS is managed by the Astrophysical Research Consortium for the Participating Institutions of the SDSS Collaboration including the Brazilian Participation Group, the Carnegie Institution for Science, Carnegie Mellon University, the Chilean Participation Group, the French Participation Group, Harvard-Smithsonian Center for Astrophysics, Instituto de Astrof\'isica de Canarias, The Johns Hopkins University, Kavli Institute for the Physics and Mathematics of the Universe (IPMU) / University of Tokyo, Lawrence Berkeley National Laboratory, Leibniz Institut f{\"u}r Astrophysik Potsdam (AIP), Max-Planck-Institut f{\"u}r Astronomie (MPIA Heidelberg), Max-Planck-Institut f{\"u}r Astrophysik (MPA Garching), Max-Planck-Institut f{\"u}r Extraterrestrische Physik (MPE), National Astronomical Observatories of China, New Mexico State University, New York University, University of Notre Dame, Observat\'orio Nacional / MCTI, The Ohio State University, Pennsylvania State University, Shanghai Astronomical Observatory, United Kingdom Participation Group, Universidad Nacional Aut\'onoma de M\'exico, University of Arizona, University of Colorado Boulder, University of Oxford, University of Portsmouth, University of Utah, University of Virginia, University of Washington, University of Wisconsin, Vanderbilt University, and Yale University.

SRT thanks Sylvie Adenis for her help improving the grammar and the style of this paper.
\end{document}